\documentstyle[12pt]{article}

\newcommand{\half}{{1\over2}}
\newcommand{\beq}{\begin{equation}}
\newcommand{\eeq}{\end{equation}}
\newcommand{\beqa}{\begin{eqnarray}}
\newcommand{\eeqa}{\end{eqnarray}}

\begin{document}
\thispagestyle{empty}
\hfill{ITP-SB-93-17}

\hfill{hep-th/9304150}
\vskip 1.5in
\begin{center}
\baselineskip24pt
{\Large  \bf Virasoro Characters from Bethe Equations for the Critical
Ferromagnetic Three-State Potts Model}
 \vskip 2cm
\baselineskip12pt
{{Srinandan Dasmahapatra,}
\\ [2mm]
{\small \em High Energy Physics, ICTP, P.O.~Box 586,
 I-34100, Trieste, Italy}}\\ [6mm]
{{Rinat Kedem, ~Barry M.~McCoy, ~and ~Ezer
Melzer~\footnote{
rinat ~or ~mccoy ~or~ melzer~@max.physics.sunysb.edu}}
\\ [2mm] {\small \em
Institute for Theoretical Physics, SUNY, Stony
Brook, NY 11794-3840, USA}}
\vskip3cm
\end{center}

\begin{abstract}
{\normalsize
We obtain new fermionic sum representations for the Virasoro
characters of the confromal field theory describing the
ferromagnetic three-state Potts spin chain.
These arise from the fermionic quasi-particle excitations derived
from the Bethe equations for the eigenvalues of the hamiltonian.
In the conformal scaling limit, the Bethe equations provide a
description of the spectrum in terms of one genuine quasi-particle,
and two ``ghost'' excitations with a limited microscopic momentum
range. This description is reflected in the structure of the
character formulas, and suggests a connection with the integrable
perturbation of dimensions (2/3,2/3)$^+$ which breaks the
$S_3$ symmetry of the conformal field theory down to ${\bf Z}_2$.
}
\end{abstract}
\baselineskip24pt
\vfill\eject
\section{Introduction}\label{intro}
\setcounter{equation}{0}
The critical three-state Potts model was found to be integrable over
20 years ago \cite{temperley71,baxter73}, and since these initial
investigations it has been the subject of many studies \cite{baxter82}.
Recently \cite{kedem93}, it was shown that the order one excitations
of the anti-ferromag\-ne\-tic three-state Potts spin chain
\cite{albertini92}, computed from the formalism of functional and
Bethe equations \cite{baxter82b}-\cite{pearce92},
can be used to construct expressions for the characters of the
conformal field theory of ${\bf Z}_4$-parafermions. Since these
equations yield excitations which obey a fermionic exclusion rule, we
call these fermionic sum representations. These character formulas
were previously obtained by Lepowsky and Primc \cite{lepowsky85} from
considerations of the representation theory of the affine Lie algebra
$A_1^{(1)}$.  The characters, which in this case are branching
functions of $(A_1^{(1)})_4/U(1)$, are the
building blocks of
the modular invariant
partition function of the conformal field theory.

Here we provide a parallel
discussion for the ferromagnetic chain, leading to fermionic sum
representations for the Virasoro characters \cite{rocha85} of the
${\bf Z}_3$-parafermionic conformal field theory which is associated
with this model \cite{dotsenko84,zamolodchikov85,cardy86a}.  These
representations, which we will now summarize, are quite different from
the ones of \cite{lepowsky85}.

\vspace{7pt}

The normalized Virasoro characters $\widehat{\chi}_\Delta
\equiv q^{\frac{1}{30}-\Delta}\chi_\Delta$ of the ${\bf Z}_3$-parafermionic
conformal field theory, with central charge $c=\frac{4}{5}$ and
conformal dimensions $\Delta=~\Delta_{r,s}~=~\frac{(6r-5s)^2-1}{120}$
{}~($r=1,2,3,4$, $s=1,3,5$), are given by~\cite{rocha85}
\beq
\widehat{\chi}_{\Delta_{r,s}}(q)~=
 ~\widehat{\chi}_{\Delta_{5-r,6-s}}(q)~= ~
 \frac{1}{(q)_{\infty}}~
  \sum_{k=-\infty}^{\infty}\left[q^{k(30k+6r-5s)}-q^{(5k+r)(6k+s)}\right]~~.
\label{chroca}
\eeq
Our result here
is that these characters can be written in the form
\beq
\widehat{\chi}_\Delta (q) =
 \sum_{{m_1,m_2,m_3=0\atop {\rm restrictions}}}^\infty
 q^{\frac{1}{4}{\bf m}
C_{A_3} {\bf m}^t - \frac{1}{2} L({\bf m})} \frac{1}{(q)_{m_1}}
\left[\begin{array}{c}\frac{1}{2} (m_1+m_3+u_2) \\ m_2 \end{array}\right]_q
\left[\begin{array}{c}\frac{1}{2} (m_2+u_3) \\ m_3 \end{array}
\right]_q\ ,
\label{gensum}
\eeq
where $(q)_0=1$, $(q)_m = \prod_{a=1}^m (1-q^a)$, the $q$-binomial
coefficient is defined for integer $m,\ n$ as
\beq
\left[\begin{array}{c}n \\m\end{array}\right]_q =
\left\{\begin{array}{cl} \frac{(q)_n}{(q)_{n-m} (q)_m} & \mbox{if
$n\geq m\geq 0$} \\ 0 & \mbox{otherwise,}\end{array}\right.
\eeq
${\bf m}=(m_1,m_2,m_3)$,
and $C_{A_3}$ is the Cartan matrix of the Lie algebra
$A_3$:
\beq
C_{A_3}=\left(\begin{array}{rrr}2 & -1 & 0 \\ -1 & 2 & -1 \\ 0 & -1 & 2
\end{array} \right).
\eeq
The restrictions on the integers $m_a$ in equation (\ref{gensum})
depend on the character in question, and are such that $m_a$ are either
even (e) or odd (o). These restrictions are listed, together with the
$u_a$ and the linear translation terms $L(\bf m)$, in table
\ref{restrictions}.  We note that for characters other than
$\widehat{\chi}_0$ and $\widehat{\chi}_3$ there is more than one
representation of the form (\ref{gensum}), and that the formulas
corresponding to lines (1)--(7),(9),(12)--(13) in the table are special
cases of the fermionic sum representations for Virasoro characters
presented already in \cite{kedem93c}.

\begin{table}
{\small
\begin{center}
\begin{tabular}{|rclllrrc|}\hline
&$\Delta$~~&$m_1$&$m_2$&$m_3$~~&$u_2$&$u_3$&~~~~$L({\bf m})$ \\ \hline
(1)  & $0$        &e&e&e&0&0&0\\ \hline
(2)  & ${2\over5}$& o&e&e&1&0&1 \\
(3)  &            &o&o&o&0&1&1 \\ \hline
(4)  & ${7\over5}$&e&e&o&1&0&3\\
(5)  &            &e&o&e&0&1&3\\ \hline
(6)  &$3$         &o&e&o&0&0&6\\ \hline
(7)  &${1\over15}$&o&e&o&2&0&$m_2$\\
(8)  &            &e&e&e&2&0&$m_2$\\
(9)  &            &e&o&o&1&1&$m_2$\\
(10) &            &o&o&e&1&1&$m_2$\\
(11) &            &$\{$e&o&o& & & \\
     &            & +o&o&e$\}$&1&$-1$&$m_1-m_3$\\ \hline
(12) &${2\over3}$ &e&e&o&1&0&$m_2+1$\\
(13) &            &o&e&e&1&0&$m_2+1$\\
(14) &            &$\{$e&o&e& & & \\
     &            &+o&o&o$\}$&0&$-1$&$m_1-m_3+1$\\ \hline
\end{tabular}
\caption{Restrictions and linear translation terms for the characters
$\widehat{\chi}_\Delta$ in equation (1.2).  Here e $\equiv$ even and o
$\equiv$ odd. Note that the characters $\widehat{\chi}_{1/15}$ and
$\widehat{\chi}_{2/3}$ have a two-term expression as well as one-term
expressions.}
\label{restrictions}
\end{center}}
\end{table}

The modular invariant partition function of the conformal field theory
associated with the three-state Potts model is written in a factorized
form in terms of these characters \cite{cardy86a}:
\beqa
(q\bar{q})^{-\frac{1}{30}}
\widehat{Z} =& [\chi_0 (q) + \chi_3(q)][\chi_0 (\bar q) + \chi_3(\bar q)]
+ [\chi_{\frac{2}{5}}(q) +\chi_{\frac{7}{5}}
(q)][\chi_{\frac{2}{5}}(\bar q) +\chi_{\frac{7}{5}} (\bar q)]&
\nonumber\\ &+
2 \chi_{\frac{1}{15}}(q)~\chi_{\frac{1}{15}}(\bar q)
+ 2\chi_{\frac{2}{3}}(q)~\chi_{\frac{2}{3}}(\bar q)~~.&
\label{mipf}
\eeqa
Here the variable $q$ (=$\bar q$) is associated with contributions
coming from right- (left-) moving excitations,
as discussed in sect.~2.

\vspace{7pt}

In this paper we construct a direct connection between the low-lying
spectrum of the spin chain hamiltonian and the conformal field theory.
We do so by computing the partition function of the spin chain in an
appropriate scaling limit (see (\ref{sclim}) below), obtaining
expressions of the form (\ref{gensum}) for the Virasoro characters.
Our starting point is the quasi-particle nature of the spectrum.

A many-body system is said to have a quasi-particle spectrum if in the
infinite-size limit the energy $E$ and momentum $P$ of the low-lying
excitations above the ground state are of the form
\beq
 E-E_{GS} ~= ~\sum_{\alpha,\ {\rm rules}} ~\sum_{j=1}^{m_\alpha}
    e_\alpha(P_j^\alpha)~,~~~~~~
 P-P_{GS} ~\equiv ~\sum_{\alpha,\ {\rm rules}} ~\sum_{j=1}^{m_\alpha}
 P_j^\alpha  \ \
  ({\rm mod}\ 2\pi)~,
\label{qpenmom}
\eeq
where $m_\alpha$ is the number of excitations of type $\alpha$ in a
given state.  The rules of composition in (\ref{qpenmom}) depend on
the model in question, and commonly include a fermionic exclusion rule
\beq
 P_j^\alpha \neq P_k^\alpha \qquad\qquad {\rm if} \quad j\neq k~,
\label{fermi}
\eeq
in which case the spectrum is said to be fermionic.

There are many cases where the excitation spectrum is gapless, {\it
i.e.}~one or more of the $e_\alpha(P^\alpha)$ vanish at some value of
the momentum, say at $P^\alpha= 0$, and
\beq
 e_\alpha(P^\alpha) ~\sim~ v_\alpha |P^\alpha| ~~~~~~{\rm for}
  ~~~~~~ P^\alpha~\sim~0~,
\label{disp}
\eeq
where $v_\alpha>0$ is the fermi velocity of the excitation of type $\alpha$.

The partition function of the quantum spin chain at temperature $T$ is
the sum over all states,
\beq
Z~=~\sum_{\{\rm states\}}  {\rm e}^{-E/k_BT}~
=~ {\rm e}^{-E_{GS}/k_BT}\sum_{\{\rm states\}}
{\rm e}^{-(E-E_{GS})/k_BT}~~ ,
\label{partfc}
\eeq
and the specific heat in the thermodynamic limit is defined by
\beq
 C(T) ~=~ -T~\frac{\partial^2 f}{\partial T^2}~~,~~~~~~~~{\rm where}~~~~
  f= -k_B T \lim_{M\to \infty} \frac{1}{M} \ln Z~.
\label{specht}
\eeq
Here $M$ is the size of the system and the temperature $T$ has some
fixed positive value.  When the spectrum is of the form (\ref{disp}),
at low temperature the specific heat is dominated by quasi-particle
states (\ref{qpenmom}) with vanishing single-particle energies and
exhibits a linear $T$ behavior. Therefore, in order to extract this
behavior it is necessary to consider only excitations of this type in
the sum over states (\ref{partfc}).  We refer to the resulting
partition function, in the limit $M\rightarrow\infty$ and with the
ground state energy factored out, as the conformal partition function.
More explicitly, the conformal field theory partition function
(\ref{mipf}) is obtained from
\beq
\widehat{Z} =
 \lim ~     {\rm e}^{E_{GS}/k_BT} Z~
\label{scaled}
\eeq
in the limit
\beq
  T \to 0~~~~~{\rm and}~~~~~M\to \infty,~~~~~
  {\rm with}~~~MT~~~{\rm fixed}.
\label{sclim}
\eeq
Using (\ref{qpenmom}) and (\ref{disp}) we see that $\widehat{Z}$ is a
function of the variable
\beq
 q~\equiv~\exp\left({\textstyle -\frac{2 \pi v}{Mk_B T}} \right)~.
\label{qdef}
\eeq

If there are no additional length scales in the problem, the $q\to 1$
behavior of $\widehat{Z}$ and the $T\rightarrow0$ limit of the
partition function in the thermodynamic limit (\ref{specht}) should
match. Indeed, the leading $q\rightarrow1$ behavior of $\widehat{Z}$
was computed in \cite{kedem93c} from the expression for the characters
(\ref{gensum}), where it was shown that the linear behavior of the
specific heat obtained in this way is the same as that obtained in the
thermodynamic limit at low temperature \cite{bazhanov89,kedem93b}.

We also remark that the connection between these two different
computations goes beyond giving just the same final result for the
value of the specific heat coefficient. In the analysis of the $q\to
1$ behavior of sums generalizing (\ref{gensum}) for characters of a
large class of conformal field theories, one encounters
\cite{kedem93c} the same equations (involving dilogarithms) which
appear in thermodynamic Bethe Ansatz analyses of the corresponding
spin chains, as well as of factorizable scattering theories that are
associated with certain integrable perturbations of the conformal
field theory in question. We will say more about the relation between
fermionic character sums and integrable perturbations in sect.~5.

\vspace{7pt}

In ~\cite{kedem93} it was shown for the anti-ferromagnetic three-state
Potts chain that the sum over low-lying excitations with a massless
dispersion relation (\ref{disp}) gives rise to the $D_4$
{}~\cite{gepner87} modular-invariant partition functions of the ${\bf
Z}_4$-parafermionic conformal field theory. In that model there are
three different excitations, all having the same linear dispersion
relation.  In contrast, the spectrum of the ferromagnetic three-state
Potts chain has a different structure. While there is only one type of
quasi-particle excitation of the kind found for the anti-ferromagnetic
case \cite{albertini92}, there are two more excitations, which do not
contribute to the energies at order one ($=M^0$) but rather determine
the degeneracy of states of the order one excitation spectrum
\cite{albertini92a}, thus affecting the thermodynamics through entropy
considerations.  In the calculation of the partition function, where
we take the energy of all excitations to have a linear dispersion
relation, this can be viewed as a statement that the momentum range of
these latter two excitations is microscopic (of order $M^{-1}$),
instead of being macroscopic (order $M^0$) as it is for the
quasi-particle excitation.

In sect.~\ref{form} we define the model and introduce the relevant
Bethe equations, as well as the order one spectrum. In sect.~\ref{q0}
we use the finite-size studies of \cite{albertini92a,klumper91} to
extend the order one analysis of the spectrum \cite{albertini92} to
order $1/M$, and study the sectors of the partition function which
give rise to the representations (1), (2), (4) and (6) in table
\ref{restrictions} for the characters
$\widehat{\chi}_0,\widehat{\chi}_3,\widehat{\chi}_{2/5}$ and
$\widehat{\chi}_{7/5}$. The sector of the partition function which
corresponds to the character $\widehat{\chi}_{1/15}$ is analyzed in
sect. \ref{q1}.  This gives a representation for
$\widehat{\chi}_{1/15}$ in terms of five sums of the form
(\ref{gensum}).  In sect.~\ref{disc} we contrast the form
(\ref{gensum}) with the result of \cite{lepowsky85}, and discuss the
relation of these different fermionic representations for the
conformal field theory characters to certain integrable massive
extensions.

\section{The gapless three-state Potts chain}\label{form}
\setcounter{equation}{0}
The gapless three-state Potts quantum spin chain of $M$ sites with
periodic boundary conditions is defined by the hamiltonian
\beq
H\ =\ \pm{ 2\over{\sqrt{3}}}~\sum_{j=1}^{M}\
\left\{X_j+ X_{j}^{\dagger}+Z_j
Z_{j+1}^{\dagger}+ Z_j^{\dagger}Z_{j+1}\right\}~,
\label{ham}
\eeq
where $Z_{M+1}=Z_1$ and for $j=1,\ldots,M$ the matrices $X_j$ and
$Z_j$ are written as a direct product of $M$ $3\times3$ matrices:
\beq
 X_{j}=I\otimes I\otimes \cdots \otimes {\underbrace X_{j^{th}}}
 \otimes \cdots \otimes I,\qquad
Z_{j}=I\otimes I \otimes \cdots \otimes {\underbrace Z_{j^{th}}}
 \otimes\cdots \otimes I~.
\eeq
Here $I$ is the identity matrix and
\beq
X = \left( \begin{array}{ccc} 0 & 0 & 1 \\ 1 & 0 & 0 \\ 0 & 1 & 0
\end{array}  \right),\qquad Z= \left(\begin{array}{ccc}1&0&0\\
0&\omega&0\\ 0&0&\omega^2\end{array}\right),\qquad \omega={\rm e}^{2 \pi i/3}.
\eeq
The hamiltonian with the $(+)\ -$ sign is referred to as the (anti-)
ferromagnetic spin chain.  The hamiltonian has a ${\bf Z}_3$
spin-rotation invariance and thus the ${\bf Z}_3$ charges $Q=0,\pm 1$
are good quantum numbers. In addition (\ref{ham}) is invariant under
complex conjugation and hence the sectors $Q=\pm 1$ have equal
eigenvalues and in $Q=0$ the eigenvalue $C=\pm 1$ of the charge
conjugation operator is a good quantum number.

The hamiltonian~(\ref{ham}) is derived from the two-dimensional
critical three-state Potts model of classical statistical mechanics.
The eigenvalues of the transfer matrix of the latter model satisfy
functional equations
\cite{albertini92b}, which, when specialized to the
hamiltonian point \cite{albertini92a}, yield equations for the
eigenvalues of the hamiltonian.  These eigenvalues are given by
\beq
E=\sum_{j=1}^{L} \cot{(i \lambda_j + \frac{\pi}{12})} - \frac{2 M}{\sqrt3}~,
{}~~~\qquad L=2(M-|Q|),\quad Q=0,\pm1,
\label{energy}
\eeq
where the rapidities $\lambda_j$ satisfy a set of equations of the form of
Bethe equations:
\beq
\left[\frac{\sinh(i \pi/12-\lambda_j)}{\sinh(i\pi/12+\lambda_j)}
\right]^{2M}
=(-1)^{M+1}  \prod_{k=1}^L\ \frac{\sinh(i \pi/3-(\lambda_j-\lambda_k))}
{\sinh(i \pi/3+(\lambda_j-\lambda_k))}~~,~~~\ \ \ j=1,\ldots,L.
\label{bae}
\eeq
The corresponding momentum, which is defined as the eigenvalue of the
translation operator, is given by
\beq{
{\rm e}^{i P } = \prod_{k=1}^L {{\sinh(\lambda_k -i\pi/12
 )}\over{\sinh(\lambda_k +i\pi/12)}}}~~.
\label{momdef}
\eeq

The solutions of the Bethe equations are sets of (possibly complex)
roots $\{\lambda_j\}$, and in the large lattice limit each root
belongs to one of five different classes \cite{albertini92a}, the
roots in each class having a fixed value of the imaginary part of
$\lambda_j$
\footnote{Note that the definition of $\lambda$ here has a
factor of $-1/2$ relative to the definition in \cite{albertini92a}.}:
\beq
\lambda_j ~~{\rm is~called}~~
\left\{\begin{array}{l}
\lambda_j^+\\  \lambda_j^-\\ \lambda_j^{2s}\\ \lambda_j^{-2s}\\
\lambda_j^{ns}
\end{array}\right\} ~~~~~{\rm if}~~~~~
\Im{\rm m} (\lambda_j) =
\left\{\begin{array}{c}
0\\ \pi/2 \\ \pm \pi/6\\ \pm \pi/3 \\ \pm \pi/4
\end{array}\right\}~.
\label{lamja}
\eeq
The last three classes of roots occur in complex conjugate pairs, and
are referred to as complex pairs. We define $m_\alpha$
(where $\alpha=+,-,2s,-2s,ns$) to be the number of roots in each class,
complex pairs being counted once.  A detailed analysis of the
equations (\ref{bae}) was performed in \cite{albertini92a}. We
summarize those results of that paper which we will use here in the
appendix.

The order one excitation spectrum obtained from
(\ref{energy})--(\ref{bae}) in the limit $M\to\infty$ was found in
\cite{albertini92}. It was shown there that for the ferromagnetic
case, the order one energy gaps can all be written in the
quasi-particle form
\beq
E-E_{GS}=\sum_{j=1}^{m_{+}} e_+(P_j^{+})~,
\label{eplus}
\eeq
where $m_+=2m_{ns}+3m_-+4m_{-2s}$ and the single-particle energy is
\beq
 e_+(P^+_j) = 6 \sin(| P^+_j |/ 2)\qquad
{}~~~0 \leq P^+_j \leq 2\pi~~,
\label{enf}
\eeq
so that the fermi velocity is $v$=3.  The momentum of a single-particle
state is expressed in terms of its rapidity $\lambda^+$ as
\beq
P ( \lambda^+) \equiv \pi + 4 \tan^{-1} (\tanh 3 \lambda^+)
{}~~({\rm mod\ } 2\pi).
\eeq

The number of states characterized by the same set $\{P_j^+\}$ (and
thus by the corresponding set of single-particle energies
$\{e_+(P_j^+)\}$) is, in the sector $Q=0$,~\cite{albertini92a}
\beq
{{m_{-}+m_{-2s}}\choose{m_{-}}}
{{2 m_{-}+2m_{-2s}+m_{ns}}\choose{m_{ns}}},
\eeq
where ${a\choose b}$ is the binomial coefficient
and $m_+=2m_{ns}+3m_-+4m_{-2s}$. This stems from the fact that the
other excitations ($ns,-2s$) carry no energy, yet states differing
only in their content of $\{ \lambda_j^\alpha\}_{\alpha=ns,-2s}$ have
to be counted individually.

In order to construct the scaled partition function (\ref{scaled}) of
the model, we extend the order one spectrum to momenta near zero. At
such momenta, the energy is
\beq
e_+(P^+) = \cases{ v P^+ & \qquad for \qquad $P^+\sim 0$ \cr
                   v (2\pi-P^+) & \qquad for \qquad $P^+\sim 2\pi$. \cr}
\label{microdisp}
\eeq
Note that there are no absolute value signs, and the momentum is no
longer defined mod $2\pi$.  This amounts to extending the order one
result (\ref{enf}) to order $1/M$; however, at this order we must
consider two additional contributions to the energy:
\begin{enumerate}
\item To order one, the excitations $ns$ and $-2s$ contribute zero
energy. However to order $1/M$ they may carry energy,
and indeed we find that  $e_\alpha(P^\alpha)=v P^\alpha$ for
$\alpha=ns,-2s$,
but with $P^\alpha$ restricted to only a microscopic range, of order
$1/M$. Here $v$ is the same as in equation (\ref{microdisp}).
\item Constant (independent of momentum or the number of excitations)
contributions of order $1/M$ to the energy must be accounted for.
These contributions, which give the conformal dimensions
$\Delta_{r,s}$, have been computed from functional equations for the
transfer matrix by
Kl\"umper and Pearce \cite{klumper91}.
\end{enumerate}

{}From equation (\ref{momdef}) and equations
(\ref{mominta}), (\ref{momintb}) and (\ref{momintc}) of the
appendix, we see that the total momentum of any state can be written as
\beq
P= \frac{2\pi}{M}\left\{\sum_{j=1}^{m_+}\bar I_j^+ +
\sum_{j=1}^{m_{-2s}}I_j^{-2s}+
\sum_{j=1}^{m_{ns}}I_j^{ns}
+ L(m_\alpha)\right\},
\eeq
where $L(m_\alpha)$ is some linear shift which depends on the sector under
consideration.  In equation (\ref{microdisp}) the energy depends
linearly on the $P_j^\alpha$, which are quantized in units of $2\pi/M$
and are directly related to the (half-) integers of the logarithmic
Bethe equations (\ref{logbae}) as:
\beq
P_j^+ = \frac{2\pi}{M} I_j^+ + \pi\equiv \frac{2\pi}{M}\bar I_j^+ ,
 \qquad
P_j^{-2s} = \frac{2\pi}{M} I_j^{-2s} , \qquad
P_j^{ns} = \frac{2\pi}{M} I_j^{ns} .
\label{momint}
\eeq
The energy can thus be expressed in terms of the
$I_j^\alpha$.

The spectrum (and so the partition function) splits into different
sectors of definite ${\bf Z}_3$ charge $Q=0,\pm 1$, and furthermore
the sector $Q=0$ splits into subsectors of parity number $C=\pm1$,
corresponding to $m_-$ being even ($C=1$) or odd ($C=-1$).  Hence we
can discuss separately each sector, which give rise to different
characters, as in the anti-ferromagnetic case~\cite{kedem93}.

\section{The characters in the sector $Q=0$}\label{q0}
\setcounter{equation}{0}
The (half-) integers in this sector are chosen from the ranges
(\ref{intrange})--(\ref{rangeq0}), and hence we see from
(\ref{momint}) that the $P_j^\alpha$ are chosen from the ranges of
spacing $2\pi/M$ with the following limits:
\beq
-\frac{2\pi}{M}\ \left[\frac{1}{2} (m_- + m_{-2s} -1)\right]\ \leq
P_j^+ \leq\ 2\pi +\frac{2\pi}{M}\
 \left[\frac{1}{2} (m_- + m_{-2s} -1)\right]\label{momrange0+}
\eeq
\beq
-\frac{2\pi}{M}\ \left[\frac{1}{2} (m_- + m_{-2s} -1)\right]\ \leq
P_j^{-2s} \leq\ \frac{2\pi}{M}\ \left[\frac{1}{2} (m_- + m_{-2s} -1)\right]
\label{momrange0-}
\eeq
\beq
-\frac{2\pi}{M}\ \left[\frac{1}{2} (2m_- +2 m_{-2s}+m_{ns} -1)\right]\ \leq
P_j^{ns} \leq\ \frac{2\pi}{M}\
\left[\frac{1}{2} (2m_- +2 m_{-2s}+m_{ns} -1)\right]~.
\label{momrange0ns}
\eeq
As is the case for the excitations of the anti-ferromagnetic chain
\cite{kedem93}, the range of
single-particle momenta for the `$+$'-excitations is macroscopic: it is
of order $2\pi$ for any finite $m_\alpha$ in the limit
$M\rightarrow\infty$.  In contrast, the ranges for $P_j^{ns}$ and
$P_j^{-2s}$ are of order $1/M$ and allow only a finite number of
states in the limit $M\rightarrow\infty$,
for given $m_\alpha$.
We refer to excitations with such microscopic momentum ranges as ``ghost''
excitations.

One expects that the partition function factorizes into right- and
left-moving contributions, as in the anti-ferromagnetic case, so that
the characters of the model are obtained by considering these
contributions separately. However, in the ferromagnetic case only the
$P_j^+$ can be considered to be right- or left-moving, where right-
(left-) movers indicates $P_j^+ \sim 0\ ~(P_j^+\sim 2\pi)$.

When taking the limit $M\rightarrow\infty$, right- (left-) movers can
be considered to lie on a semi-infinite range, since the range for
$P^+$ is macroscopic, allowing for an infinite number of momentum
states. Therefore, we rewrite the momentum range for right-movers in
this limit as
\beq
-\frac{2\pi}{M}\left[\frac{1}{2}(m_-+m_{-2s}-1)\right]\leq
P_j^+<\infty~~~~~~~~~~\hbox{\rm for~right-movers},
\label{momrange0++}
\eeq
replacing equation (\ref{momrange0+}) above.  For the left-movers it
is convenient to replace $P^+$ by $P^+-2\pi$, so that the momentum
range in the $M\to\infty$ limit is
\beq
-\infty < P_j^+ \leq \frac{2\pi}{M}\left[\frac{1}{2}(m_-+m_{-2s}-1)\right]
 ~~~~~~~~~~\hbox{\rm for~left-movers},
\label{momrange0++l}
\eeq
and the dispersion relation (\ref{microdisp}) now reads
\beq
e_+(P^+) = \cases{ v P^+ & \qquad for right-movers, \cr
                   -v P^+ & \qquad for left-movers. \cr}
\label{microdispnew}
\eeq

There are four characters corresponding to the $Q=0$ sector
(since there is a symmetry between right- and left-movers,
below we restrict our attention to the right-movers):
\begin{enumerate}
\item The vacuum character $\widehat{\chi}_0$,
which corresponds to the sector of the partition
function with only right-movers
and positive parity, $C=+1$;
\item The character $\widehat{\chi}_3$,
which corresponds to the sector with only right-movers and negative
parity, $C=-1$;
\item The character $\widehat{\chi}_{2/5}$, which has one
left-mover and the rest right-movers, with $C=+1$, and
\item $\widehat{\chi}_{7/5}$, which has   only   one
left-mover and $C=-1$.
\end{enumerate}
We will discuss in detail the construction of the partition function
in the sector corresponding to $\widehat{\chi}_0$, item 1 above,
and then outline the computation of the other characters.

\subsection{Construction of the character $\widehat{\chi}_0$}
The sector of the partition function (\ref{partfc}) which has only
right-movers and $Q=0,\ C=+1$ is computed as follows. The excitation
energy is simply the sum over the individual excitations near
$P^\alpha\sim0$
\beq
E-E_{GS} =   v \left\{\sum_{j=1}^{m_+} P_j^+ +
\sum_{j=1}^{m{_-2s}} P_j^{-2s} + \sum_{j=1}^{m_{ns}}
P_j^{ns}\right\}~.
\eeq
The partition function is the sum over all right-moving excitations
with momentum ranges (\ref{momrange0-})--(\ref{momrange0++}), subject
to the fermionic exclusion rule (\ref{fermi}), and the restriction
that $m_-$ be even.  In table \ref{sec01} we show the lowest energy
states of in this sector.  The general expression for the partition
function in this sector is
\beq
\widehat{Z}_0
= \sum_{\{{\rm states}\}} {\rm e}^{-vP/k_BT}
= \sum_{\{I_j^\alpha\}}
q^{\left(\sum_j \bar I_j^+ + \sum_j I_j^{-2s} + \sum_j I_j^{ns} \right)} ~~,
\label{part0}
\eeq
with $q$ defined as in equation (\ref{qdef}).  Here the $I_j^\alpha$
are restricted as in equation (\ref{intrange}) in the appendix with
$m_-$ even.  As in (\ref{momrange0++}), in the limit $M\to\infty$ we
have for right-moving `$+$'-excitations
\beq
-{1\over2}(m_-+m_{-2s}-1)\leq \bar I_j^+<\infty\ .\label{inf}
\eeq

The restrictions on the integers are implemented by using two integer
partitions, $Q_m(N;n)$ and $Q_m(N)\equiv Q_m(N;\infty)$, where
$Q_m(N;n)$ is the number of partitions of $N\geq0$ into $m$ distinct
non-negative integers each less than or equal to $n$.  The partition
function (\ref{part0}) subject to the restrictions
(\ref{momrange0++}),(\ref{momrange0-}),(\ref{momrange0ns}) then
becomes:
\beqa
\widehat{Z}_0  =
\sum_{{m_-,m_{-2s},m_{ns}=0 \atop m_-\ {\rm even}}
\atop m_+ = 2 m_{ns}+3m_- + 4 m_{-2s}}^\infty
\sum_{N_+,N_{-2s},N_{ns}=0}^\infty
Q_{m_+}(N_+)~ q^{N_+-\frac{1}{2}m_+(m_{-2s}+m_{-}-1)}
{}~~\hbox{\hskip.5in}
  \nonumber\\
  \times ~~
Q_{m_{-2s}}(N_{-2s}; m_{-2s}+m_- - 1)
 ~q^{N_{-2s}-\frac{1}{2}m_{-2s}(m_{-2s}+m_{-}-1)}  ~~
\nonumber\\
\times  ~~ Q_{m_{ns}}(N_{ns};m_{ns}+2m_{-2s}+2m_--1 )
{}~q^{N_{ns}-\frac{1}{2}m_{ns}(m_{ns}+2m_{-2s}+2m_{-}-1)}~~.\nonumber\\
\label{z0a}
\eeqa
The exponents of $q$ above are essentially the total momenta of each
type of excitation, {\it i.e.} the sums over the integers
$N_\alpha=\sum_j I_j^\alpha$.  The partitions count the number of times
$q^{\sum_\alpha N_\alpha}$ occurs in the partition function, which is
the number of ways $N_\alpha$ can be divided between $m_\alpha$
fermionic excitations.

The sum (\ref{z0a}) can be re-expressed using the identity
\cite{stanley72,andrews76}
\beq
\sum_{N=0}^\infty Q_m(N;n)q^N = q^{m(m-1)/2} \left[\begin{array}{c}
n+1 \\ m \end{array} \right]_q\ ,
\label{part2}
\eeq
which, when $n\to\infty$, reduces to
\beq
\sum_{N=0}^\infty Q_m(N) q^N = \frac{q^{m(m-1)/2}}{(q)_m}~~.
\label{part1}
\eeq
Using these identities, (\ref{z0a}) becomes
\beqa
\widehat{Z}_0 =
 \sum_{{m_-,m_{-2s},m_{ns}=0\atop m_+ = 2 m_{ns} +3 m_- + 4 m_{-2s}}\atop
m_- \ {\rm even}}^\infty
q^{-\frac{1}{2}(m_++m_{-2s})(m_{-2s}+m_- -1)
-\frac{1}{2}m_{ns}(m_{ns}+2m_{-2s}+2m_- -1)}\nonumber\\
 \times ~
q^{\frac{1}{2}m_+(m_+-1)+\frac{1}{2}m_{-2s}(m_{-2s}-1)+
\frac{1}{2}m_{ns}(m_{ns}-1)}\nonumber\\
 \times ~ \frac{1}{(q)_{m_+}}
\left[\begin{array}{c} m_{-2s}+m_- \\ m_{-2s} \end{array} \right]_q
\left[\begin{array}{c} m_{ns}+2m_{-2s}+2m_- \\ m_{ns} \end{array}
\right]_q~.
\label{z0}
\eeqa
The form of the sum may be further simplified by changing variables to
\beq
m_1 = m_+~,\qquad
m_2 = 2m_- + 2m_{-2s}~,\qquad
m_3 = m_-~,
\label{malphas}
\eeq
which results in the expression (\ref{gensum}) with the restriction
that all $m_a$ are even and $L({\bf m})=0,\ u_a=0$. This is the
expression listed on line (1) of table \ref{restrictions}.

This expression is quite different in form from the one given in
(\ref{chroca}). Nevertheless we find that
\beq
\widehat{Z}_0~ =~\widehat{\chi}_0 ~~.
\eeq
This has been verified as an equality between the series expansions of
the two expressions to order $q^{200}$, using Mathematica.

\subsection{Construction of $\widehat{\chi}_3$}
In table \ref{sec02} we present the lowest energy states of the sector
$C=-1$, where all $m_+$ are right-movers. The calculation of the
partition function $\widehat Z_3$ is identical to that of the last
section, except that now $m_-$ is odd in equations (\ref{z0a}) and
(\ref{z0}).  Using series expansions, we verify that the resulting
expression $\widehat Z_3$ is equal to $q^3 \widehat{\chi}_3$ to order
$q^{200}$.  With the change of variables (\ref{malphas}), this results
in the expression on line (6) of table \ref{restrictions}.

\subsection{Construction of $\widehat{\chi}_{2/5}$ and $\widehat{\chi}_{7/5}$}

The characters $\widehat{\chi}_{2/5}$ and $\widehat{\chi}_{7/5}$ occur
when one of the $m_+$ is a left-mover, and all the rest are
right-movers. This amounts to setting $m_+ = 2 m_{ns} + 3 m_- +4
m_{-2s} -1$ in the partition sum (\ref{z0a}) and (\ref{z0}).  Also,
there is an additive term to the momentum of the form $\frac{\pi
v}{M}(m_{-}+m_{ns}-1)$, which is the lowest energy state of the single
left-moving `$+$'-excitation allowed by equation (\ref{momrange0++l}).
The character $\widehat\chi_{2/5}$ occurs for $C=+1$, {\it i.e.} $m_-$
even, and the character $\widehat\chi_{7/5}$ occurs for $C=-1$, $m_-$
odd. We tabulate the lowest energy states for these two sectors in
tables \ref{sec03} and
\ref{sec04}.  The expression for the resulting partition functions is
\beqa
\widehat{Z}_{2/5~(7/5)}  =
\sum_{{m_+,m_{-2s},m_{ns}=0 \atop m_+ =
2 m_{ns}+3m_- + 4 m_{-2s}-1}\atop m_-\
{\rm even~ (odd)}}
^\infty  \sum_{N_\alpha=0}^\infty
q^{\frac{1}{2}(m_-+m_{-2s}-1)}
 Q_{m_+}(N_+)~ q^{N_+-\frac{1}{2}m_+(m_{-2s}+m_{-}-1)}  \nonumber\\
\times ~
Q_{m_{-2s}}(N_{-2s}; m_{-2s}+m_- - 1)
 ~q^{N_{-2s}-\frac{1}{2}m_{-2s}(m_{-2s}+m_{-}-1)}
\nonumber\\
\times ~  Q_{m_{ns}}(N_{ns};m_{ns}+2m_{-2s}+2m_--1 )
{}~q^{N_{ns}-\frac{1}{2}m_{ns}(m_{ns}+2m_{-2s}+2m_{-}-1)}~.\nonumber\\
\label{z5}
\eeqa
Using the identities (\ref{part2})--(\ref{part1}) and the change of variables
(\ref{malphas}),
$\widehat Z_{2/5}$ and $q^{-1}\widehat Z_{7/5}$ are brought to the form
(\ref{gensum}) with the restrictions listed on
lines (2) and (4) of table \ref{restrictions}, {\it i.e.} we find that
\beq
\widehat Z_{2/5} ~=~ \widehat\chi_{2/5}~, \qquad ~~
\widehat Z_{7/5} ~=~ q \widehat\chi_{7/5}~.
\eeq
The other expressions for the two characters $\widehat\chi_{2/5}$ and
$\widehat\chi_{7/5}$, corresponding to lines (3) and (5) of that
table, are conjectured forms. Using a power series expansion, all
these forms were shown to be equal to the corresponding expressions
(\ref{chroca}) for $\widehat\chi_{2/5}$ and $\widehat\chi_{7/5}$, to
order $q^{200}$.

\section{The Sector $Q=1$}\label{q1}
\setcounter{equation}{0}
The analysis of this sector is more involved than for the $Q=0$
sector, since (see appendix) there are five different sub-sectors to
be considered, where the integers range over different intervals.
Each of these sub-sectors gives rise to a separate sum in the sector
of the partition function corresponding to right-moving excitations.
The resulting five sums add together to form the Virasoro character
$\widehat\chi_{1/15}$, as we now describe in more detail.

The momentum ranges are given by the integer ranges (\ref{intrange})
and (\ref{q1-1})--(\ref{q1-5}) in the appendix, and the relations
(\ref{mominta}), (\ref{momintb}) and (\ref{momintc}) of the total momentum
to the integers.  We will present the computation for each sub-sector
separately, where for each sub-sector all `$+$'-excitations are
right-movers.
\begin{enumerate}
\item $m_- -m_{++}= +1$: Here
we see from equation (\ref{mp1}) that $m_+ = 2m_{ns} + 3 m_{-} +
4m_{-2s}-2$, and the lowest energy states are shown in table
\ref{sec11}, where the integer range is that of equation (\ref{q1-1}).
The partition sum starts with $m_-=1$, since $m_->m_{++}\geq0$:
\beqa
\widehat{Z}_{1/15}^{(1)} =
\sum_{m_- =1}^{\infty}\sum_{m_{ns},m_{-2s}=0\atop
m_+ = 2 m_{ns} + 3 m_- + 4 m_{-2s} -2}^\infty
q^{\frac{1}{2}m_+(m_+ -1) + \frac{1}{2}m_+(3-m_--m_{-2s})}\frac{1}{(q)_{m_+}}
\nonumber\\
\times~ q^{\frac{1}{2}m_{-2s}(m_{-2s} -1) + \frac{1}{2}m_{-2s}(3-m_--m_{-2s})}
\left[\begin{array}{c} m_{-2s}+m_--2 \\ m_{-2s} \end{array} \right]_q
\nonumber\\ \times ~
q^{\frac{1}{2}m_{ns}(m_{ns} -1) + \frac{1}{2}m_{ns}(3-2m_--2m_{-2s}-m_{ns})}
\left[\begin{array}{c} m_{ns}+2m_{-2s}+2m_--2 \\ m_{ns} \end{array}\right]_q.
\nonumber\\
\eeqa
\item $m_--m_{++}=-1$: From the sum rule (\ref{mp1}) we see that $m_+ = 2m_{ns}
+ 3 m_- + 4 m_{-2s} + 2$, and the integer range is given by
(\ref{q1-2}). The
lowest energy states
are shown in
table \ref{sec12}, and the general expression for the partition function is:
\beqa
\widehat{Z}_{1/15}^{(2)} =
\sum_{m_-,m_{ns},m_{-2s}=0\atop
m_+ = 2 m_{ns} + 3 m_- + 4 m_{-2s} +2}^\infty
q^{\frac{1}{2}m_+(m_+ -1) + \frac{1}{2}m_+(1-m_--m_{-2s})}\frac{1}{(q)_{m_+}}
\nonumber\\
\times ~q^{\frac{1}{2}m_{-2s}(m_{-2s} -1) + \frac{1}{2}m_{-2s}(1-m_--m_{-2s})}
\left[\begin{array}{c} m_{-2s}+m_- \\ m_{-2s} \end{array} \right]_q
\nonumber\\ \times~
q^{\frac{1}{2}m_{ns}(m_{ns} -1) + \frac{1}{2}m_{ns}(-1-2m_--2m_{-2s}-m_{ns})}
\left[\begin{array}{c} m_{ns}+2m_{-2s}+2m_-+2 \\ m_{ns} \end{array}\right]_q.
\nonumber\\
\eeqa
\item $m_- = m_{++} = 0$: Since $m_-=0$, there are no
$-2s$ excitations.  The relevant momentum range is obtained from the
integer range (\ref{q1-3}), and the lowest energy states are listed in
table \ref{sec13}.  The general expression for the partition function
is a simple sum over $ns$ excitations, with $m_+ = 2 m_{ns}$:
\beq
\widehat{Z}_{1/15}^{(3)} =
\sum_{m_{ns}=0\atop m_+=2m_{ns}}^\infty \frac{q^{{m_+}(m_++1)/2}}
{(q)_{m_+}} =
\sum_{m_{ns}=0}^\infty \frac{q^{m_{ns}(2m_{ns}+1)}}{(q)_{2m_{ns}}}~~.
\label{s1}
\eeq
\item $m_- = m_{++} \neq 0$: There are two sub-sectors with this
characteristic, corresponding to the integer ranges (\ref{q1-4}) and
(\ref{q1-5}).  These integer ranges are asymmetric, so there is a
shift term in the total momentum, as shown in (\ref{momintb}) and
(\ref{momintc}), of the form $\mp\frac{2\pi v}{M}
({1\over2}m_{ns}+m_-+m_{-2s})$.  For both of these sectors $m_+ = 2
m_{ns} + 3m_- + 4 m_{-2s}$. The lowest energy states are listed in
tables
\ref{sec14} and \ref{sec15}.  The sums take the forms, for the integer range
(\ref{q1-4}):
\beqa
\widehat{Z}_{1/15}^{(4)} =
\sum_{m_- =1}^{\infty}\sum_{m_{ns},m_{-2s}=0\atop
m_+ = 2 m_{ns} + 3 m_- + 4 m_{-2s}}^\infty
q^{-(m_- + m_{-2s} + {1\over2}m_{ns})}
q^{\frac{1}{2}m_+(m_+ -1) + \frac{1}{2}m_+
 ({3}-m_--m_{-2s})}\frac{1}{(q)_{m_+}}
\nonumber\\
\times~ q^{\frac{1}{2}m_{-2s}(m_{-2s} -1) + \frac{1}{2}m_{-2s}(
{1}-m_--m_{-2s})}
\left[\begin{array}{c} m_{-2s}+m_- \\ m_{-2s} \end{array} \right]_q
\nonumber\\ \times~
q^{\frac{1}{2}m_{ns}(m_{ns} -1) + \frac{1}{2}m_{ns}(2m_--2m_{-2s}-m_{ns})}
\left[\begin{array}{c} m_{ns}+2m_{-2s}+2m_- \\ m_{ns}
\end{array}\right]_q~,
\nonumber\\
\eeqa
and for the integer range (\ref{q1-5}):
\beqa
\widehat{Z}_{1/15}^{(5)} =
\sum_{m_- =1}^{\infty}\sum_{m_{ns},m_{-2s}=0\atop
m_+ = 2 m_{ns} + 3 m_- + 4 m_{-2s}}^\infty
q^{(m_- + m_{-2s} + {1\over2}m_{ns})}
q^{\frac{1}{2}m_+(m_+ -1) + \frac{1}{2}m_+
 ({1}-m_--m_{-2s})}\frac{1}{(q)_{m_+}}
\nonumber\\
\times~ q^{\frac{1}{2}m_{-2s}(m_{-2s} -1) + \frac{1}{2}m_{-2s}(
{3}-m_--m_{-2s})}
\left[\begin{array}{c} m_{-2s}+m_- \\ m_{-2s} \end{array} \right]_q
\nonumber\\ \times~
q^{\frac{1}{2}m_{ns}(m_{ns} -1) + \frac{1}{2}m_{ns}(2-2m_--2m_{-2s}-m_{ns})}
\left[\begin{array}{c} m_{ns}+2m_{-2s}+2m_- \\ m_{ns}
\end{array}\right]_q~.
\nonumber\\
\eeqa
\end{enumerate}
Finally, we find that
\beq
 \sum_{a=1}^5 \widehat{Z}_{1/15}^{(a)}~ =~ \widehat\chi_{1/15}~.
\eeq
This is a five-sum expression for the character $\widehat\chi_{1/15}$, where
each summand can be expressed in the form (\ref{gensum}).  In addition to
this form, one can find the forms listed in table \ref{restrictions}
for the character $\widehat\chi_{1/15}$. Again, although all these forms are
quite different from that of (\ref{chroca}), they have been shown to
be equal to order $q^{200}$.

It remains to consider the character $\widehat{\chi}_{2/3}$. Here,
however, no analysis corresponding to the above five-term sum form is
availible. The conjectured forms on lines (12)--(14) of table 1 have
been verified to order $q^{200}$.

\section{Discussion}\label{disc}
\setcounter{equation}{0}

The forms of the expressions (\ref{chroca}) and (\ref{gensum}) for the
characters of the ferromagnetic three-state Potts conformal field
theory deserve to be called ``different'', even though the expressions
are equal.  The question thus arises as to what is meant by the word
different, how many different forms there are, and what their
significance is.  We know of at least four different forms for the
characters of the three-state Potts.  One is the Rocha-Caridi form
(\ref{chroca}), the second is the form of Kac and Peterson
\cite{kac84} and Jimbo and Miwa \cite{jimbo84},
the third is that of Lepowsky and Primc \cite{lepowsky85}, and the
fourth is the form (\ref{gensum}).  Each of these forms is
sufficiently different to warrant a separate discussion.

\begin{enumerate}
\item
The expression (\ref{chroca}) for the Virasoro characters, which are
\cite{goddard86} branching functions of the coset
{}~$\frac{(A_1^{(1)})_3 \times (A_1^{(1)})_1}{(A_1^{(1)})_4}$, is what
we refer to as a bosonic sum representation. This stems from the
presence of the factor $(q)_\infty^{-1}$, which represents a bosonic
partition function and can be understood in terms of the
Feigin-Fuchs-Felder construction \cite{feigin83,felder89} of the
Virasoro minimal series \cite{belavin84} ${\cal M}(p,p')$ to which the
three-state Potts conformal field theory belongs, being ${\cal
M}(5,6)$ in this notation.
\item
The second form is also a bosonic expression which can be obtained by
viewing this conformal field theory as that of ${\bf
Z}_3$-parafermions~\cite{zamolodchikov85}, where the characters of the
corresponding ${\bf Z}_3$-parafermionic algebra are~\cite{gepner87}
the branching functions of the coset ~$\frac{(A_1^{(1)})_3}{U(1)}$.
Another description of the same conformal field theory is as a minimal
model with respect to the $W_3$ algebra~\cite{fateev87}, where the
corresponding coset construction is ~$\frac{(A_2^{(1)})_1 \times
(A_2^{(1)})_1}{(A_2^{(1)})_2}$.  The latter construction is related by
level-rank duality~\cite{altschuler90} to
{}~$\frac{(A_1^{(1)})_3}{U(1)}$, and the branching functions are in fact
the same. They are given by the Hecke indefinite forms
of~\cite{kac84,jimbo84} (or alternative but very similar sum
representations of~\cite{distler90})
\beqa
q^{\frac{1}{30}}b_m^l(q) = \frac{q^{h_m^l}}{(q)_{\infty}^2}
\Bigg[\Bigg(\sum_{s\geq0}\sum_{n\geq0}{}-\sum_{s<0}\sum_{n<0}\Bigg)
(-1)^s q^{s(s+1)/2+(l+1)n+(l+m)s/2+5(n+s)n}\nonumber\\
+ \Bigg( \sum_{s>0}\sum_{n\geq0}-\sum_{s\leq0}\sum_{n<0}\Bigg)
(-1)^s q^{s(s+1)/2+(l+1)n+(l-m)s/2+5(n+s)n}\Bigg]~,
\label{blm}
\eeqa
where the   $h^l_m$ are
\beq
 h_m^{l}=\frac{l(l+2)}{20}-\frac{m^2}{12}~~.
\label{hlm}
\eeq
Here $l=0,1,2$, ~$l-m$ is even, and the formulas are valid for
$|m| \leq l$ while for $|m|>l$ one uses the symmetries
\beq
 b_m^l=b_{-m}^l=b_{m+6}^l=b_{3-m}^{3-l}~.
\label{symm}
\eeq
The partition function (\ref{mipf}) is expressed as a diagonal
bilinear form in terms of the $b^l_m$, through
\beq
\chi_0 + \chi_3 = b^0_0~,~~~ \chi_{2/5}+\chi_{7/5} = b^2_0~,
{}~~~\chi_{1/15} = b^2_2~,~~~ \chi_{2/3}=b^0_2~.
\label{connection}
\eeq
Note that two of the $b^l_m$ split into a sum of a pair of Virasoro
characters, corresponding to a more refined splitting of the spectrum
of the hamiltonian into various sectors.  Also, the expressions
(\ref{chroca}) have only one factor of $(q)_\infty^{-1}$ while the
ones in (\ref{blm}) have two. Thus whereas (\ref{chroca}) can be said
to be based on one boson, (\ref{blm}) is based on two bosons.
\item
The third form is a fermionic sum representation for the branching
functions $b^l_m$ which was obtained by Lepowsky and Primc
\cite{lepowsky85}:
\beq
q^{1/30} b_{2Q-l}^l(q) = q^{\frac{l(2-l)}{16}}
\sum_{m_1,m_2=0\atop m_1-m_2\equiv Q\ ({\rm mod~3})}^\infty
\frac{q^{{\bf m}C^{-1}_{A_2}{\bf m}^t+L_l({\bf m})}}{(q)_{m_1}(q)_{m_2}}~~,
\label{LepPrim}
\eeq
where $C_{A_2} = \left(\begin{array}{rr}2&-1\\-1&2\end{array}\right)$
is the Cartan matrix of the Lie algebra $A_2$, and
$L_0({\bf m})=0$, $L_1({\bf m})=(2m_1+m_2)/3$,
$L_2({\bf m})=(m_1+2m_2)/3$.
This expression can be interpreted \cite{kedem93a}
in terms of two  $C$-conjugate fermionic quasi-particles
carrying ${\bf Z}_3$ charges $\pm1$, both having macroscopic momentum
ranges.
For example, for $l=0$ these ranges are
\beq
\frac{2\pi}{M}\left[ \frac{1}{2} + \frac{1}{2}\left(\frac{m_1+2m_2}{3}\right)
\right] \leq P_j^1 < \infty~,\qquad
\frac{2\pi}{M}\left[ \frac{1}{2} + \frac{1}{2}\left(\frac{2m_1+m_2}{3}\right)
\right] \leq P_j^2 <\infty~,
\label{LPmom}
\eeq
where the $P_j^\alpha$ ~($j=1,\ldots,m_\alpha$)
are chosen from a grid with spacing $2\pi/M$.
\item
The fourth form is the fermionic sum representation (\ref{gensum})
which has one genuine quasi-particle with a macroscopic momentum range
and two ``ghost'' quasi-particles, whose momenta are limited to a
microscopic range, {\it e.g.}~equations
(\ref{momrange0+})--(\ref{momrange0ns}).
\end{enumerate}

The existence of different fermionic sum representations for
characters is closely related to the fact that one conformal field
theory may have several integrable perturbations, characterized by the
conformal dimensions $(\Delta, \overline{\Delta})$ of certain
perturbing relevant operators.  In \cite{kedem93a} this observation
was made in connection with the representations of the critical Ising
characters as related to either (i) the coset $(A_1^{(1)})_1\times
(A_1^{(1)})_1\over (A_1^{(1)})_2$ where the character formulas are
written in terms of a single quasi-particle and the associated
perturbation is by the $(1/2,1/2)$ operator, or (ii) the coset
$(E_8^{(1)})_1\times (E_8^{(1)})_1\over (E_8^{(1)})_2$ which has eight
quasi-particles and is associated with the $(1/16,1/16)$ perturbation
\cite{zamolodchikov89}. In each case the perturbation can be thought
of as giving masses to the fermionic quasi-particles.

A similar discussion can be given for the two different fermionic
representations (\ref{LepPrim}) and (\ref{gensum}) of the critical
three-state Potts model.  Consider first (\ref{LepPrim}), which was
interpreted as having two fermionic quasi-particles of ${\bf Z}_3$
charge $\pm1$. This is to be compared with the $(2/5,2/5)$
$S_3$-symmetric perturbation of the three-state Potts conformal field
theory, which was argued in
\cite{zamolodchikov88} to be integrable and to have a spectrum which
consists of a ${\bf Z}_3$-doublet of massive particles, whose
scattering is described by the factorizible $S$-matrix found in
\cite{koberle79}. Here the perturbation  can again be thought of as
giving mass to the two fermionic quasi-particles. This perturbation is
also to be compared with the massive $r=5$ RSOS model (or hard squares
with diagonal interactions) \cite{baxter80}, whose spectrum on the
lattice \cite{baxter82b} in regime II consists of two excitations
with ${\bf Z}_3$ charge $\pm1$.

In the same spirit it is natural to associate the fermionic sum
representation (\ref{gensum}) with the $C$-even $(2/3,2/3)^+$
perturbation.  This subleading magnetic perturbation breaks the $S_3$
symmetry down to ${\bf Z}_2$.  The related statistical mechanics
system is the $N$=3 model of Kashiwara and Miwa \cite{kashiwara86}
(also given as the $D_4$ model of Pasquier \cite{pasquier87}, obtained
from the $r$=6 RSOS model of Andrews, Baxter and Forrester
\cite{andrews84} by an orbifold construction \cite{fendley89}).  In
the notation of \cite{pasquier87,andrews84} the lattice models are to
be considered in the regimes III-IV.  The perturbed conformal field
theory is the $p$=5 case of the
$(\Delta_{1,3},\Delta_{1,3})$-perturbed minimal models ${\cal
M}(p,p+1)$, which have been discussed in
\cite{zamolodchikov89}\cite{zamolodchikov87}-\cite{klassen91} where
it is seen that the sign of the coupling to the perturbing operator
leads to qualitatively very different effects.

In the case of negative coupling constant (using the conventions of
\cite{zamolodchikov89}) the perturbed theory becomes massive. This is
to be compared to the massive regime III of the models of
\cite{kashiwara86,pasquier87,andrews84},
where excitation energies have been computed in \cite{bazhanov90a}. We
interpret this direction of the perturabtion as giving mass to the
quasi-particle $m_1$ of (\ref{gensum}).

The more interesting case is the one where the coupling constant is
positive.  Now the integrable perturbed conformal field theory remains
massless \cite{zamolodchikov89,alzam91}, even though scale invariance
is broken, and flows
\cite{fateev91,ravanini92,klassen91,cardy88} from the three-state
Potts conformal field theory of central charge 4/5 to the conformal
field theory of the tricritical Ising model of central charge 7/10.
This suggests an interpretation in terms of the representation
(\ref{gensum}), where we note that under the restriction to the sector
where there are no ``ghost'' excitations of type $m_3$ the fermionic
representations for the three-state Potts characters reduce to
fermionic representations \cite{kedem93c} for the characters of the
tricritical Ising model.  Specifically, restricting the summation in
(\ref{gensum}) by setting $m_3=0$, the formulas corresponding to lines
(1), (2), (5), (8), (10) and (13) of table 1 reduce to expressions for the
$c=7/10$ Virasoro characters $\widehat{\chi}_\Delta$ with
$\Delta$=0, 7/16, 3/2, 3/80, 1/10 and 3/5, respectively.  The crucial
point making this possible is the fact that (four times) the quadratic
form in the fermionic sum representations of the $c=7/10$ characters
is the Cartan matrix of $A_2$, which is precisely the minor, obtained
by omitting the last row and column, of the quadratic form $C_{A_3}$
in (\ref{gensum}).  More generally, we find from \cite{kedem93c} that
the fermionic form of the characters of the unitary minimal model
${\cal M}(p,p+1)$ with one quasi-particle and $p-3$ ``ghosts'' reduces
to character formulas for ${\cal M}(p-1,p)$ when the last ghost is
omitted, the corresponding massless flows being the ones discussed in
\cite{zamolodchikov87,ludwig87,alzam91}.

\vspace{1cm}

\noindent
{\bf Acknowledgements:} We would like to thank Dr.~G.~Albertini
and Prof.~V.V.~Bazhanov   for
fruitful
discussions.  The work of RK and BMM is partially supported by the
NSF, grant DMR-9106648, and that of EM by the NSF grant 91-08054.

\vfill\eject
\begin{appendix}
\section{Appendix: Logarithmic Bethe Equations}\label{Fapp}
\setcounter{equation}{0}
We recall here some results \cite{albertini92a} concerning the
classification of the solutions of the Bethe equations corresponding
to the eigenvalues of the hamiltonian (\ref{ham}).

Not all the roots $\lambda_j^\alpha$ (cf.~equation (\ref{lamja})) in a
given solution of the Bethe equations (\ref{bae}) are independent of
one another, and in order to discuss the relations between them we
introduce the logarithmic Bethe equations.  By taking the logarithm of
the Bethe equations (\ref{bae}), we can classify the sets
$\{\lambda_j^\alpha\}$ more easily.  Doing this introduces integers or
half-integers associated with the choice of branch of the logarithm.
The equations for the complex pairs are first multiplied together.
After taking the logarithm, we obtain five sets of equations, one for
each class of roots, referred to as the logarithmic Bethe equations:
\beq f_\alpha
\frac{2\pi }{M}I_j^\alpha=
t_\alpha(\lambda_j^\alpha)-
\frac{1}{M}\sum_{\beta=\pm,\pm2s,ns}\ \ \sum_{k=1}^{m_\beta}
\Theta_{\alpha\beta}(\lambda_j^\alpha-\lambda_k^\beta)~, ~~~~~~
\alpha\in\{+,-,2s,-2s,ns\},
\label{logbae}
\eeq
where $f_{\alpha}=2$ for $\alpha=ns$
and is 1 otherwise
\footnote{Note that the factor $f_{ns}$
was not present in the definitions of \cite{albertini92a}. This amounts to a
 redefinition of the integers $I^{ns}$ discussed there.},
and
where the functions $\Theta_{\alpha\beta}$ and $t_\alpha$ are defined
as follows. Let $
s_\alpha (\lambda) \equiv {\sinh(i\alpha-\lambda)/\sinh(i\alpha+\lambda)},
$
then
\beqa
t_\alpha(\lambda_j^\alpha) &=&\cases{
 -2i \ln(\pm s_{\pi/12}(\lambda_j^\pm)) &  $\alpha=\pm$\cr
 -2i \ln(s_{\pi/12}(\lambda_j^\alpha)s_{\pi/12}(\lambda_j^{\alpha*}))~~~~~&
 $\alpha=\pm 2s,ns, $\cr}\label{defs1}\nonumber\\ \\
\Theta_{\alpha\beta}(\lambda_j^\alpha-\lambda_k^\beta) &=&
\cases{
 -i \ln(\epsilon_{\alpha,\beta}\ s_{\pi/3}(\lambda_j^\alpha-\lambda_k^\beta)) &
 $\alpha,\beta=\pm$\cr
 -i  \ln (\epsilon_{\alpha,\beta}\ s_{\pi/3}(\lambda_j^\alpha -
\lambda_k^\beta) s_{\pi/3}(\lambda_j^\alpha-\lambda_k^{\beta *})) &
 $\alpha=\pm, \beta=\pm2s, ns$ \cr
 -i  \ln (\epsilon_{\alpha,\beta}\ s_{\pi/3}(\lambda_j^\alpha -
\lambda_k^\beta) s_{\pi/3}(\lambda_j^{\alpha *}-\lambda_k^{\beta})) &
 $\alpha=\pm2s, ns, \beta=\pm$ \cr
 -i \ln (\epsilon_{\alpha,\beta}\
 s_{\pi/3}(\lambda_j^\alpha - \lambda_k^\beta)
 s_{\pi/3}(\lambda_j^\alpha - \lambda_k^{\beta *}) & \cr
 \hbox{\hskip.45in} \times~s_{\pi/3}(\lambda_j^{\alpha *} - \lambda_k^\beta)
 s_{\pi/3}(\lambda_j^{\alpha *} - \lambda_k^{\beta *}) )~~ &
$\alpha,\beta = \pm2s, ns$, \cr}\nonumber\\
\label{defs2}
\eeqa
where the symmetric tensor $\epsilon_{\alpha,\beta}$ is defined by
$\epsilon_{+,-}
=\epsilon_{-,2s}=\epsilon_{+,-2s}=\epsilon_{2s,2s}=\epsilon_{-2s,-2s}
= -1$ and the other $\epsilon_{\alpha,\beta}$ are 1.
$\epsilon_{\alpha,\beta}$ is chosen so that
$\Theta_{\alpha\beta}(\lambda^\alpha-\lambda^\beta)=0$ when $\Re{\rm
e} \lambda^\alpha = \Re{\rm e} \lambda^\beta$.  All logarithms in
(\ref{defs1}) and (\ref{defs2}) are chosen such that $-\pi < \Im{\rm
m}\ln{z} < \pi$.  Each set of (half-) integers $\{I_j^\alpha\}$
uniquely specifies a set of roots $\{\lambda_j^\alpha\}$. Note that
the sets contain {\em either} integers {\em or} half-integers,
depending on $m_\alpha$.

For the sector $Q=0$, there is a restriction on the number $m_+$ of
the form
\beq
m_+=2n_{ns}+3m_-+4m_{-2s}~.
\label{sumq0}
\eeq
In addition, the total number of roots is $2M$ (see equation
(\ref{energy})), so that
\beq
M =m_{2s}+2m_{ns}+3m_{-2s}+2m_-~.
\label{tot0}
\eeq
For the sector $Q=\pm1$, we define the number $m_{++}$ which has the
property that
$m_- - m_{++} = 0,\pm1.$
For this sector we have the sum rule
\beq
m_+=2n_{ns}+m_- + 2 m_{++} + 4m_{-2s}~,\label{mp1}
\eeq
and since the total number of roots is $2(M-1)$ (see equation
(\ref{energy})), we have
\beq
M-1 =m_{2s}+2m_{ns}+3m_{-2s}+m_- + m_{++} ~.
\eeq

The (half-) integers in equation (\ref{logbae}) are not all
independent, as the set $\{I_j^{2s}\}$ and $\{I_j^{-}\}$ are
completely determined from the sets $\{I_j^{+}\}$ and $\{I_j^{-2s}\}$,
respectively.  The ground state of the ferromagnetic chain consists of
a sea of $2s$-excitations, that is the integers $\{I_j^{2s}\}$ fill a
symmetric interval about zero, and all other sets of integers are null
sets.  Therefore, for convenience,
we take the sets $\{I_j^{+}\}$, $\{I_j^{-2s}\}$ and
$\{I_j^{ns}\}$ to be the independent sets  in
discussing the ferromagnetic case. Those (half-) integers are then
freely chosen from the intervals
\beqa
-\frac{1}{2} \left[ M + m_- + m_{-2s} - a_\ell^{(1)}\right]\
\leq& I_j^+& \leq\ +
\frac{1}{2} \left[ M + m_- + m_{-2s} - a_r^{(1)} \right]\nonumber\\
-\frac{1}{2} \left[  m_- + m_{-2s} - a_r^{(1)}\right]\
\leq& I_j^{-2s}& \leq\ +
\frac{1}{2} \left[  m_- + m_{-2s} - a_\ell^{(1)} \right]\nonumber\\
-\frac{1}{2} \left[2m_- + 2m_{-2s}+m_{ns} - a_\ell^{(2)}\right]\
\leq& I_j^{ns}& \leq\ +
\ \frac{1}{2} \left[2 m_- +2 m_{-2s}+m_{ns} - a_r^{(2)} \right]
\label{intrange}
\eeqa
with a fermionic exclusion rule:
$I_j^\alpha \neq I_k^\alpha~~{\rm for}~j\neq k$.
The numbers $a_\ell$ and $a_r$ depend on the sector in question.
For the $Q=0$ sector,
\beq
a_\ell^{(1)}=a_r^{(1)}=
a_\ell^{(2)}=a_r^{(2)}=1 \quad~~~~ {\rm for}~~~~ Q=0~.\label{rangeq0}
\eeq
 In the $Q=\pm1$
sectors, there are five separate sub-sectors to be considered, depending
on the value of $m_{++}$ introduced above:
\beqa
{\rm For}~~& m_- - m_{++} = +1~:~~ & a_\ell^{(1)} = a_r^{(1)} = 3,\
a_\ell^{(2)} = a_r^{(2)} = 3
\label{q1-1}\\
{\rm For}~~& m_- - m_{++} = -1~:~~ & a_\ell^{(1)} = a_r^{(1)} = 1,\
a_\ell^{(2)} = a_r^{(2)} = -1
\label{q1-2}\\
{\rm For}~~& m_- = m_{++} = 0~:~~ & a_\ell^{(1)} = a_r^{(1)} = 2,\
a_\ell^{(2)} = a_r^{(2)} = 1 \
\label{q1-3} \\
{\rm For}~~& m_- = m_{++} \neq 0~:~~ & a_\ell^{(1)} = 3,\ a_r^{(1)} = 1,\
a_\ell^{(2)} = 0,\  a_r^{(2)} = 2
\label{q1-4}  \\
{\rm For}~~& m_- = m_{++} \neq 0~:~~ & a_\ell^{(1)} = 1,\ a_r^{(1)} = 3,\
a_\ell^{(2)} = 2,\ a_r^{(2)} = 0~.\label{q1-5}
\eeqa
The last two sectors correspond to two degenerate sets of energy
eigenvalues.

The total momentum of each state is determined from equation
(\ref{momdef}), and can be expressed in terms of $\{I_j^\alpha\}$
using the logarithmic Bethe equations (\ref{logbae}). Taking the
logarithm of equation (\ref{momdef}) and using the definitions
(\ref{defs1}), the total momentum can be written as
\beqa
P &\equiv&\half \sum_{j=1}^{m_+} t_+(\lambda_j^+)+
{1\over2}\sum_{j=1}^{m_-}\left( t_-(\lambda_j^-)+2\pi\right)+
{1\over2}\sum_{j=1}^{m_{2s}} t_{2s}(\lambda_j^{2s}) \nonumber\\ & &
\hbox{\hskip1.1in}  +
{1\over2}\sum_{j=1}^{m_{-2s}} t_{-2s}(\lambda_j^{-2s})+
\half \sum_{j=1}^{m_{ns}} t_{ns}(\lambda_j^{ns})~~({\rm mod}~2\pi)~.
\eeqa
We sum the logarithmic Bethe equations (\ref{logbae}) over $j$ and
$\alpha$. The sum over the functions $\Theta_{\alpha\beta}$ vanishes
since they are odd functions. We are left with a sum over the
integers:
\beq
P \equiv \frac{2\pi}{M} \left(\half\sum_{j=1}^{m_+}I_j^+ +
{1\over2}\sum_{j=1}^{m_-}\left( I_j^-+M\right) +
{1\over2}\sum_{j=1}^{m_{2s}} I_j^{2s}+
{1\over2}\sum_{j=1}^{m_{-2s}} I_j^{-2s}+
\sum_{j=1}^{m_{ns}} I_j^{ns}\right)~({\rm mod~}2\pi)~.
\eeq
In order to express the momentum in terms of three independent sets of
integers, we note that for the sector $Q=0$, as well as for the
sectors corresponding to equations (\ref{q1-1})--(\ref{q1-3}), where
the (half-) integers are chosen from a symmetric interval about zero,
the two sets of (half-) integers $\{I_j^+\}$ and $\{-I_j^{2s}\}$ fill
this interval, and similarly for the sets $\{I_j^-\}$ and
$\{-I_j^{-2s}\}$.  Therefore,
\beq
\sum_{j=1}^{m_{\pm2s}} I_j^{\pm2s} - \sum_{j=1}^{m_\pm} I_j^{\pm}=0~,
\eeq
and the total momentum of a state may be written (using $m_+\equiv
m_-~({\rm mod~}2)$) as
\beq
P \equiv \frac{2\pi}{M} \left(\sum_{j=1}^{m_+}\bar I_j^{+} +
                         \sum_{j=1}^{m_{-2s}}I_j^{-2s} +
                         \sum_{j=1}^{m_{ns}}I_j^{ns}\right)~({\rm mod~}2\pi)~,
\label{mominta}
\eeq
where $\bar I_j^+ = I_j^++M/2$.

However, for the sectors corresponding to equations
(\ref{q1-4})--(\ref{q1-5}) there is an additional term involved, since
the integer ranges are not symmetric about zero, and there is an
offset between the sets $\{I_j^{\pm}\}$ and $\{-I_j^{\pm2s}\}$. In
fact, for the sector (\ref{q1-4}) the following relation between
the integers holds \cite{albertini92a}:
\beq
I_j^{2s\ h} = - I_j^{+} + \frac{1}{2}~, \qquad
I_j^{-\ h} = - I_j^{-2s} - \frac{1}{2}~, \label{off1}
\eeq
where the superscript $h$ refers to ``holes'', namely the (half-) integers
missing from the set $\{I_j^\alpha\}$. The number of $2s$-holes is
$m_+$, and the number of `$-$'-holes is $m_{-2s}$.
The ranges of integers are chosen such that:
\beq
\sum_{j=1}^{m_{2s}}I_j^{2s} + \sum_{j=1}^{m_+} I_j^{2s\ h} = 0~,\label{zer1}
\eeq
that is the  $I_j^{2s}$ are chosen from a symmetric range.
This is not the case for the  $I_j^-$, which are chosen from
the range:
\beq
-\half (m_-+m_{-2s}) \leq I_j^- \leq \half (m_- + m_{-2s}-2)~,
\eeq
so that
\beq
\sum_{j=1}^{m_-} I_j^{-} + \sum_{j=1}^{m_{-2s}}I_j^{-\ h} =
-\frac{1}{2} (m_- + m_{-2s})~.\label{zer2}
\eeq
Putting equations (\ref{off1})--(\ref{off2}) together, we find that for
this sector
\beq
P \equiv \frac{2\pi}{M} \left(\sum_{j=1}^{m_+}\bar I_j^{+} +
                         \sum_{j=1}^{m_{-2s}}I_j^{-2s} +
                         \sum_{j=1}^{m_{ns}}I_j^{ns} -
\left(\frac{1}{2} m_{ns} + m_- + m_{-2s}\right)\right)~({\rm mod~}2\pi)~.
\label{momintb}
\eeq

For the sector corresponding to equation (\ref{q1-5}) we have
\beq
I_j^{2s\ h} = - I_j^{+} - \frac{1}{2}~, \qquad
I_j^{-\ h} = - I_j^{-2s} + \frac{1}{2}~. \label{off2}
\eeq
Equation (\ref{zer1}) still holds, but the range of $I_j^-$ is now
such that
\beq
\sum_{j=1}^{m_-} I_j^{-} + \sum_{j=1}^{m_{-2s}}I_j^{-\ h} =
\frac{1}{2} (m_- + m_{-2s})~.\label{zer3}
\eeq
Therefore the total momentum in this sector is found to be
\beq
P \equiv \frac{2\pi}{M} \left(\sum_{j=1}^{m_+}\bar I_j^{+} +
                         \sum_{j=1}^{m_{-2s}}I_j^{-2s} +
                         \sum_{j=1}^{m_{ns}}I_j^{ns} +
\left(\frac{1}{2} m_{ns} + m_- + m_{-2s}\right)\right)~({\rm mod~}2\pi).
\label{momintc}
\eeq

\end{appendix}
\vfill\eject

%%%TABLES
\baselineskip12pt
\tabcolsep4pt
\begin{table}
\begin{center}
{\footnotesize
\caption{The first terms of the partition function in the
sector $Q=0$ and $C=1$, where $m_-$ is even,
and $m_+=2 m_{ns}+3 m_- + 4 m_{-2s}$. The momentum ranges are
given in equations (\protect{\ref{momrange0-}})--(\protect{\ref{momrange0++}}).
The sum of the momenta in the square brackets
gives the total momentum, and thus  the power of $q$,
listed on the left. $N$ is the number of states with
given $m_\alpha$ and fixed total momentum,
whose overall number
is listed on the right. These are the coefficients of $q^n$ in the
power expansion of $\widehat \chi_0$.}\vspace{12pt}
\begin{tabular}{|l|cccc|c|c|l|r|r|} \hline
{Order} & $m_+$ & $m_{-2s}$ & $m_-$ & $m_{ns}$ & $P_{min}^{+,-2s}$
& $P_{min}^{ns}$ & $[P^{ns};P^{-2s};P^{+}]$ (Units
of $\frac{\pi}{M}$) & {$N$} & {Tot} \\ \hline \hline
$q^0$ & 0 & 0 & 0 & 0 & $-$     & $-$ & [$-$; $-$; $-$] & 1 & 1 \\ \hline
$q^2$ & 2 & 0 & 0 & 1 & $\pi/M$ & 0 & [0; $-$; 1,3] & 1 & 1 \\ \hline
$q^3$ & 2 & 0 & 0 & 1 & $\pi/M$ & 0 & [0; $-$; 1,5] & 1 & 1 \\ \hline
$q^4$ & 2 & 0 & 0 & 1 & $\pi/M$ & 0 & [0; $-$; 1,7],[0; $-$; 3,5]
 & 2 & 2 \\ \hline
$q^5$ & 2 & 0 & 0 & 1 & $\pi/M$ & 0 & [0; $-$; 1,9],[0; $-$; 3,7]
 & 2 & 2 \\ \hline
$q^6$ & 2 & 0 & 0 & 1 & $\pi/M$ & 0 & [0; $-$; 1,11],[0; $-$; 3,9],
[0; $-$; 5,7]& 3 &  \\
   & 4 & 1 & 0 & 0 & 0 & $-$ & [$-$; 0; 0,2,4,6] & 1 & 4 \\ \hline
$q^7$ & 2 & 0 & 0 & 1 & $\pi/M$ & 0 & [0; $-$; 1,13],[0; $-$; 3,11],
[0; $-$; 5,9]& 3 &  \\
   & 4 & 1 & 0 & 0 & 0 & $-$ & [$-$; 0; 0,2,4,8] & 1 & 4 \\ \hline
$q^8$ & 2 & 0 & 0 & 1 & $\pi/M$ & 0 & [0; $-$; 1,15],[0; $-$; 3,13]
&  &  \\
      &   &   &   &   &         &   & [0; $-$; 5,11],[0; $-$; 7,9] & 4 & \\
   & 4 & 1 & 0 & 0 & 0 & $-$ & [$-$; 0; 0,2,4,10],[$-$; 0; 0,2,6,8] & 2 &
\\
   &4  & 0 & 0 & 2 & $\pi/M$ & $-\pi/M$ & [$-1$,1; $-$; 1,3,5,7] & 1 & 7
\\ \hline
$q^9$ & 2 & 0 & 0 & 1 & $\pi/M$ & 0 & [0; $-$; 1,17],[0; $-$; 3,15]
&  &  \\
      &   &   &   &   &         &   & [0; $-$; 5,13],[0; $-$; 7,11] & 4 & \\
   & 4 & 1 & 0 & 0 & 0 & $-$ & [$-$; 0; 0,2,4,12],[$-$; 0; 0,2,6,10] &  &
\\
   &   &   &   &   &   &     & [$-$; 0; 0,4,6,8] & 3 & \\
   & 4 & 0 & 0 & 2 & $\pi/M$ & $-\pi/M$  & [$-1$,1; $-$; 1,3,5,9] & 1 & 8 \\
\hline
$q^{10}$ & 2 & 0 & 0 & 1 & $\pi/M$ & 0 & [0; $-$; 1,19],[0; $-$; 3,17]
&  &  \\
      &   &   &   &   &         &   &[0; $-$; 5,15], [0; $-$; 7,13],
[0; $-$; 9,11] & 5 & \\
   & 4 & 1 & 0 & 0 & 0 & $-$ & [$-$; 0; 0,2,4,14],[$-$; 0; 0,2,6,12] &  &
\\
   &   &   &   &   &   &     & [$-$; 0; 0,4,6,10],[$-$; 0; 0,2,8,10] &  & \\
   &   &   &   &   &   &     & [$-$; 0; 2,4,6,8] & 5 & \\
   & 4 & 0 & 0 & 2 & $\pi/M$ & $-\pi/M$  & [$-1$,1; $-$; 1,3,5,11],
[$-1$,1; $-$; 1,3,7,9] & 2 & 12 \\ \hline
$q^{11}$ & 2 & 0 & 0 & 1 & $\pi/M$ & 0 & [0; $-$; 1,21],[0; $-$; 3,19]
&  &  \\
      &   &   &   &   &         &   &[0; $-$; 5,17], [0; $-$; 7,15],
[0; $-$; 9,13] & 5 & \\
   & 4 & 1 & 0 & 0 & 0 & $-$ & [$-$; 0; 0,2,4,16],[$-$; 0; 0,2,6,14] &  &\\
   &   &   &   &   &   &     & [$-$; 0; 0,2,8,12],[$-$; 0; 0,4,6,12] &  & \\
   &   &   &   &   &   &     & [$-$; 0; 0,4,8,10],[$-$; 0; 2,4,6,10] & 6 & \\
   & 4 & 0 & 0 & 2 & $\pi/M$ & $-\pi/M$  & [$-1$,1; $-$; 1,3,5,13],
[$-1$,1; $-$; 1,3,7,11] &   &  \\
   &   &   &   &   &         &           & [$-1$,1; $-$; 1,5,7,9] & 3 & 14 \\
\hline
$q^{12}$ & 2 & 0 & 0 & 1 & $\pi/M$ & 0 & [0; $-$; 1,23],[0; $-$;
3,21],[0; $-$; 5,19]&  &  \\
      &   &   &   &   &         &   &[0; $-$; 7,17], [0; $-$; 9,15],
[0; $-$; 11,13] & 6 & \\
   & 4 & 1 & 0 & 0 & 0 & $-$ & [$-$; 0; 0,2,4,18],[$-$; 0; 0,2,6,16] &  &\\
   &   &   &   &   &   &     & [$-$; 0; 0,2,8,14],[$-$; 0; 0,2,10,12] &  & \\
   &   &   &   &   &   &     & [$-$; 0; 0,4,6,14],[$-$; 0; 2,4,6,12] &  & \\
   &   &   &   &   &   &     & [$-$; 0; 0,4,8,12],[$-$; 0; 2,4,8,10] &  & \\
   &   &   &   &   &   &     & [$-$; 0; 0,6,8,10] & 9 & \\
   & 4 & 0 & 0 & 2 & $\pi/M$ & $-\pi/M$  & [$-1$,1; $-$; 1,3,5,15],
[$-1$,1; $-$; 1,3,7,13] &   &  \\
   &   &   &   &   &         &           & [$-1$,1; $-$; 1,5,7,11],
   [$-1$,1; $-$; 1,3,9,11] &  &  \\
   &   &   &   &   &         &           & [$-1$,1; $-$; 3,5,7,9]& 5 &  \\
   & 6 & 0 & 2 & 0 & $-\pi/M$ & $-$ & [$-$; $-$; $-1$,1,3,5,7,9] &
   1 & 21 \\ \hline
\end{tabular}
\label{sec01}
}
\end{center}
\end{table}

\begin{table}
\begin{center}
{\footnotesize
\caption{The first few terms in the partition function in the sector
$Q=0$ and $C=-1$, corresponding to $m_-$ odd, and $m_+=2 m_{ns}+3 m_-
+ 4 m_{-2s}$.  The momentum ranges are the same as in table
\protect{\ref{sec01}}. The total number of states on the right corresponds
to the first few terms in the expansion of $q^3\widehat \chi_3$. }
\vspace{12pt}
\begin{tabular}{|l|cccc|c|c|l|r|r|} \hline
{Order} & $m_+$ & $m_{-2s}$ & $m_-$ & $m_{ns}$ & $P_{min}^{+,-2s}$
& $P_{min}^{ns}$ & $[P^{ns};P^{-2s};P^{+}]$ (Units
of $\frac{\pi}{M}$) & {$N$} & {Tot} \\ \hline\hline
$q^3$ & 3 & 0 & 1 & 0 & 0 & $-$ & [$-$; $-$; 0+2+4] & 1 & 1 \\ \hline
$q^4$ & 3 & 0 & 1 & 0 & 0 & $-$ & [$-$; $-$; 0+2+6] & 1 & 1 \\ \hline
$q^5$ & 3 & 0 & 1 & 0 & 0 & $-$ & [$-$; $-$; 0+2+8],
 [$-$; $-$; 0+4+6] & 2 & 2 \\ \hline
$q^6$ & 3 & 0 & 1 & 0 & 0 & $-$ & [$-$; $-$; 0+2+10],[$-$; $-$; 0+4+8] & & \\
      &   &   &   &   &   &     & [$-$; $-$; 2+4+6] & 3 & 3 \\ \hline
$q^7$ & 3 & 0 & 1 & 0 & 0 & $-$ & [$-$; $-$; 0+2+12],
 [$-$; $-$; 0+4+10] &  &  \\
      &   &   &   &   &   &     & [$-$; $-$; 0+6+8],
 [$-$; $-$; 2+4+8]    & 4 & 4\\ \hline
$q^8$ & 3 & 0 & 1 & 0 & 0 & $-$ & [$-$; $-$; 0+2+14],
[$-$; $-$; 0+4+12] &  &  \\
      &   &   &   &   &   &     & [$-$; $-$; 0+6+10],
 [$-$; $-$; 2+4+10] & & \\
      &   &   &   &   &   &     & [$-$; $-$; 2+6+8]   & 5 & 5\\ \hline
$q^9$ & 3 & 0 & 1 & 0 & 0 & $-$ & [$-$; $-$; 0+2+16],
 [$-$; $-$; 0+4+14] &  &  \\
      &   &   &   &   &   &     & [$-$; $-$; 2+4+12],
 [$-$; $-$; 0+6+12] & & \\
      &   &   &   &   &   &     & [$-$; $-$; 2+6+10],
 [$-$; $-$; 4+6+8]    &  &\\
      &   &   &   &   &   &     & [$-$; $-$; 0+8+10]   & 7 & \\
      & 5 & 0 & 1 & 1 & 0 & $-2\pi/M$ & [$-2$; $-$; 0+2+4+6+8] & 1 & 8 \\
\hline
$q^{10}$ & 3 & 0 & 1 & 0 & 0 & $-$ & [$-$; $-$; 0+2+18],
 [$-$; $-$; 0+4+16] &  &  \\
      &   &   &   &   &   &     & [$-$; $-$; 2+4+14],
 [$-$; $-$; 0+6+14] & & \\
      &   &   &   &   &   &     & [$-$; $-$; 2+6+12],
 [$-$; $-$; 4+6+10]   & &\\
      &   &   &   &   &   &     & [$-$; $-$; 0+8+12],
 [$-$; $-$; 2+8+10]   & 8 & \\
      & 5 & 0 & 1 & 1 & 0 & $-2\pi/M$ & [$-2$; $-$; 0+2+4+6+10] &  &  \\
      &   &   &   &   &   &           & [0 ; $-$; 0+2+4+6+8] & 2 & 10 \\
\hline
$q^{11}$ & 3 & 0 & 1 & 0 & 0 & $-$ & [$-$; $-$; 0+2+20],
 [$-$; $-$; 0+4+18] &  &  \\
      &   &   &   &   &   &     & [$-$; $-$; 2+4+16],
 [$-$; $-$; 0+6+16] & & \\
      &   &   &   &   &   &     & [$-$; $-$; 2+6+14],
 [$-$; $-$; 4+6+12]    & &\\
      &   &   &   &   &   &     & [$-$; $-$; 0+8+14],
 [$-$; $-$; 2+8+12]    &  &\\
      &   &   &   &   &   &     & [$-$; $-$; 4+8+10],
 [$-$; $-$; 0+10+12]    & 10 & \\
      & 5 & 0 & 1 & 1 & 0 & $-2\pi/M$ & [$-2$; $-$; 0+2+4+6+12] &  &  \\
      &   &   &   &   &   &           & [$-2$ ; $-$; 0+2+4+8+10] &  &  \\
      &   &   &   &   &   &           & [0 ; $-$; 0+2+4+6+10] &  &  \\
      &   &   &   &   &   &           & [2 ; $-$; 0+2+4+6+8] &4  &14  \\
\hline
$q^{12}$ & 3 & 0 & 1 & 0 & 0 & $-$ & [$-$; $-$; 0+2+22],
 [$-$; $-$; 0+4+20] &  &  \\
      &   &   &   &   &   &     & [$-$; $-$; 2+4+18],
 [$-$; $-$; 0+6+18] & & \\
      &   &   &   &   &   &     & [$-$; $-$; 2+6+16],
 [$-$; $-$; 4+6+14]   & &\\
      &   &   &   &   &   &     & [$-$; $-$; 0+8+16],
 [$-$; $-$; 2+8+14]   &  &\\
      &   &   &   &   &   &     & [$-$; $-$; 4+8+12],
 [$-$; $-$; 0+10+14]   &  &\\
      &   &   &   &   &   &     & [$-$; $-$; 2+10+12],
 [$-$; $-$; 6+8+10]   &12  & \\
      & 5 & 0 & 1 & 1 & 0 & $-2\pi/M$ & [$-2$; $-$; 0+2+4+6+14] &  &  \\
      &   &   &   &   &   &           & [$-2$ ; $-$; 0+2+4+8+12] &  &  \\
      &   &   &   &   &   &           & [$-2$ ; $-$; 0+2+6+8+10] &  &  \\
      &   &   &   &   &   &           & [0 ; $-$; 0+2+4+6+12] &  &  \\
      &   &   &   &   &   &           & [0 ; $-$; 0+2+4+8+10] &  &  \\
      &   &   &   &   &   &           & [2 ; $-$; 0+2+4+6+10] & 6 & 18 \\
\hline
\end{tabular}\label{sec02}
}
\end{center}
\end{table}
\begin{table}
\begin{center}
{\footnotesize
\caption{ The first few terms for the sector of the partition
function corresponding to $Q=0$ and $C=1$, where one of the
`$+$'-excitations is a left-mover, and the rest are right-movers. This
corresponds to $m_-$ even and $m_+=2 m_{ns}+3 m_- + 4 m_{-2s}-1 $.
The momentum ranges are the same as in table
\protect{\ref{sec01}},
and there is an additional term (``shift'') in the momentum of
$\frac{\pi}{M}(m_-+m_{-2s} - 1)$ which is the momentum of the
left-mover.  The coefficients on the right correspond to the expansion
of $\widehat \chi_{2/5}$. }\label{sec03}
\vspace{12pt}
\begin{tabular}{|l|cccc|c|c|l|c|r|r|} \hline
{Order} & $m_+$ & $m_{-2s}$ & $m_-$ & $m_{ns}$ & $P_{min}^{+,-2s}$
& $P_{min}^{ns}$ & $[P^{ns};P^{-2s};P^{+}]$ (Units
of $\frac{\pi}{M}$) & {Shift} &{$N$} & {Tot} \\ \hline\hline
$q^0$ & 1 & 0 & 0 & 1 & $\pi/M$ & 0 & [0; $-$; 1] & $-\pi/M$ & 1 & 1\\ \hline
$q^1$ & 1 & 0 & 0 & 1 & $\pi/M$ & 0 & [0; $-$; 3] & $-\pi/M$ & 1 & 1\\ \hline
$q^2$ & 1 & 0 & 0 & 1 & $\pi/M$ & 0 & [0; $-$; 5] & $-\pi/M$ & 1 & 1\\ \hline
$q^3$ & 1 & 0 & 0 & 1 & $\pi/M$ & 0 & [0; $-$; 7] & $-\pi/M$ &1  & \\
      & 3 & 1 & 0 & 0 & 0       &$-$& [$-$; 0; 0,2,4] & 0 & 1 & 2 \\ \hline
$q^4$ & 1 & 0 & 0 & 1 & $\pi/M$ & 0 & [0; $-$; 9] & $-\pi/M$ &1  & \\
      & 3 & 1 & 0 & 0 & 0       &$-$& [$-$; 0; 0,2,6] & 0 & 1 &  \\
      & 3 & 0 & 0 & 2 & $\pi/M$ & $-\pi/M$ & [$-1$,1; $-$; 1,3,5] &
            $-\pi/M$ & 1 & 3 \\ \hline
$q^5$ & 1 & 0 & 0 & 1 & $\pi/M$ & 0 & [0; $-$; 11] & $-\pi/M$ & 1 & \\
      & 3 & 1 & 0 & 0 & 0  &$-$& [$-$; 0; 0,2,8],[$-$; 0; 0,4,6] & 0 & 2 &  \\
      & 3 & 0 & 0 & 2 & $\pi/M$ & $-\pi/M$ & [$-1$,1; $-$; 1,3,7] &
            $-\pi/M$ & 1 & 4 \\ \hline
$q^6$ & 1 & 0 & 0 & 1 & $\pi/M$ & 0 & [0; $-$; 13] & $-\pi/M$ & 1 & \\
      & 3 & 1 & 0 & 0 & 0       &$-$& [$-$; 0; 0,2,10],[$-$; 0; 0,4,8] & & &\\
      &   &   &   &   &         &   & [$-$; 0; 2,4,6] &0 &3 &\\
      & 3 & 0 & 0 & 2 &$\pi/M$ & $-\pi/M$ & [$-1$,1; $-$; 1,3,9],
      [$-1$,1; $-$; 1,5,7] &        $-\pi/M$ & 2 & 6 \\ \hline
$q^7$ & 1 & 0 & 0 & 1 & $\pi/M$ & 0 & [0; $-$; 15] & $-\pi/M$ & 1 & \\
      & 3 & 1 & 0 & 0 & 0      &$-$& [$-$; 0; 0,2,12],[$-$; 0; 0,4,10] & & &\\
      &   &   &   &   &        &   & [$-$; 0; 2,4,8],[$-$; 0; 0,6,8] &0 &4 &\\
      & 3 & 0 & 0 & 2 & $\pi/M$&$-\pi/M$ & [$-1$,1; $-$; 1,3,11],
     [$-$1,1; $-$; 1,5,9] &    &  &  \\
      &   &   &   &   &         &          & [$-1$,1; $-$; 3,5,7] &
            $-\pi/M$ & 3 &8  \\ \hline
$q^8$ & 1 & 0 & 0 & 1 & $\pi/M$ & 0 & [0; $-$; 17] & $-\pi/M$ & 1 & \\
      & 3 & 1 & 0 & 0 & 0    &$-$& [$-$; 0; 0,2,14],[$-$; 0; 0,4,12] & & &\\
      &   &   &   &   &      &   & [$-$; 0; 2,4,10],[$-$; 0; 0,6,10] & & &\\
      &   &   &   &   &       &   & [$-$; 0; 2,6,8] &0 &5 &\\
      & 3 & 0 & 0 & 2&$\pi/M$& $-\pi/M$ & [$-1$,1; $-$; 1,3,13],
   [$-$1,1; $-$; 1,5,11] &   &  &  \\
      &   &   &   &   &         &        & [$-1$,1; $-$; 3,5,9],
         [$-1$,1; $-$; 1,7,9] &    $-\pi/M$ & 4 &  \\
      & 5 & 0 & 2 & 0 & $-\pi/M$ & $-$ & [$-$; $-$; $-$1,1,3,5,7] & $\pi/M$ &
                                       1 & 11 \\ \hline
$q^9$ & 1 & 0 & 0 & 1 & $\pi/M$ & 0 & [0; $-$; 19] & $-\pi/M$ &1  & \\
      & 3 & 1 & 0 & 0 & 0     &$-$& [$-$; 0; 0,2,16],[$-$; 0; 0,4,14] & & &\\
      &   &   &   &   &        &   & [$-$; 0; 2,4,12],[$-$; 0; 0,6,12] & & &\\
      &   &   &   &   &        &   & [$-$; 0; 2,6,10],[$-$; 0; 4,6,8] & & &\\
      &   &   &   &   &         &   & [$-$; 0; 0,8,10], &0 &7 &\\
      & 3 & 0 & 0 & 2&$\pi/M$& $-\pi/M$ & [$-1$,1; $-$; 1,3,15],
            [$-1$,1; $-$; 1,5,13] &   &  &  \\
      &   &   &   &   &       & & [$-1$,1; $-$; 3,5,11],[$-1$,1; $-$; 1,7,11] &
                     &  &  \\
      &   &   &   &   &         &      & [$-1$,1; $-$; 3,7,9] &
            $-\pi/M$ & 5 &  \\
      & 5 & 0 & 2 & 0 & $-\pi/M$ & $-$ & [$-$; $-$; $-1$,1,3,5,9] & $\pi/M$ &
                                       1 &  \\
      & 5 & 1 & 0 & 1 & 0 & $-2\pi/M$ & [$-$2; 0; 0,2,4,6,8] & 0 & 1 & 15
                   \\ \hline
\end{tabular}}
\end{center}
\end{table}

\begin{table}
\begin{center}
{\footnotesize
\caption{The first few terms in the partition function in the sector
$Q=0$, $C=-1$ where one `$+$'-excitation is left-moving and all the
rest are right-movers. The shift and $m_+$ are as in table
\protect{\ref{sec03}}, and the momentum ranges are as in table
\protect{\ref{sec01}}. The coefficients on the right correspond to the
expansion of $q\widehat \chi_{7/5}$. }\label{sec04}
\vspace{12pt}
\begin{tabular}{|l|cccc|c|c|l|c|r|r|} \hline
{Order} & $m_+$ & $m_{-2s}$ & $m_-$ & $m_{ns}$ & $P_{min}^{+,-2s}$
& $P_{min}^{ns}$ & $[P^{ns};P^{-2s};P^{+}]$ (Units
of $\frac{\pi}{M}$) & {Shift} &{$N$} & {Tot} \\ \hline\hline
$q^1$ & 2 & 0 & 1 & 0 & 0 & $-$ & [$-$; $-$; 0,2] & 0 & 1 & 1 \\ \hline
$q^2$ & 2 & 0 & 1 & 0 & 0 & $-$ & [$-$; $-$; 0,4] & 0 & 1 & 1 \\ \hline
$q^3$ & 2 & 0 & 1 & 0 & 0 & $-$ & [$-$; $-$; 0,6],[$-$; $-$; 2,4]
 & 0 & 2 & 2 \\ \hline
$q^4$ & 2 & 0 & 1 & 0 & 0 & $-$ & [$-$; $-$; 0,8],[$-$; $-$; 2,6]
  & 0 & 2 & 2 \\ \hline
$q^5$ & 2 & 0 & 1 & 0 & 0 & $-$ & [$-$; $-$; 0,10],[$-$; $-$; 2,8],
 [$-$; $-$; 4,6]  & 0 & 3 &  \\
      & 4 & 0 & 1 & 1 & 0 & $-2\pi/M$ & [$-2$; $-$; 0,2,4,6] & 0 & 1 & 4
\\ \hline
$q^6$ & 2 & 0 & 1 & 0 & 0 & $-$ & [$-$; $-$; 0,12],[$-$; $-$; 2,10],
 [$-$; $-$; 4,8]  & 0 & 3 &  \\
      & 4 & 0 & 1 & 1 & 0 & $-2\pi/M$ & [$-2$; $-$; 0,2,4,8],
                                        [0; $-$;  0,2,4,6] & 0 & 2 & 5
\\ \hline
$q^7$ & 2 & 0 & 1 & 0 & 0 & $-$ & [$-$; $-$; 0,14],[$-$; $-$; 2,12],
 [$-$; $-$; 4,10]        &  &  &  \\
      &   &   &   &   &   &     & [$-$; $-$; 6,8] & 0 & 4 & \\
      & 4 & 0 & 1 & 1 & 0 & $-2\pi/M$ & [$-2$; $-$; 0,2,4,10],
                                        [$-2$; $-$; 0,2,6,8] & & & \\
      &   &   &   &   &   &           & [0; $-$;  0,2,4,8],
                                        [2; $-$;  0,2,4,6] & 0 & 4 & 8\\
 \hline
$q^8$ & 2 & 0 & 1 & 0 & 0 & $-$ & [$-$; $-$; 0,16],[$-$; $-$; 2,14],
 [$-$; $-$; 4,12]   &  &  &  \\
      &   &   &   &   &   &     & [$-$; $-$; 6,10] & 0 & 4 & \\
      & 4 & 0 & 1 & 1 & 0 & $-2\pi/M$ & [$-2$; $-$; 0,2,4,12],
                                        [$-2$; $-$; 0,2,6,10] & & & \\
      &   &   &   &   &   &           & [$-2$; $-$; 0,4,6,8],
                                        [0; $-$;  0,2,4,10] & & & \\
      &   &   &   &   &   &           & [0; $-$;  0,2,6,8],
                                        [2; $-$;  0,2,4,8] & 0 & 6 & 10\\
 \hline
$q^9$ & 2 & 0 & 1 & 0 & 0 & $-$ & [$-$; $-$; 0,18],[$-$; $-$; 2,16],
[$-$; $-$; 4,14]   &  &  &  \\
      &   &   &   &   &   &     & [$-$; $-$; 6,12],[$-$; $-$; 8,10]
 & 0 & 5 & \\
      & 4 & 0 & 1 & 1 & 0 & $-2\pi/M$ & [$-2$; $-$; 0,2,4,14],
                                        [$-2$; $-$; 0,2,6,12] & & & \\
      &   &   &   &   &   &           & [$-2$; $-$; 0,4,6,10],
                                        [$-2$; $-$; 2,4,6,8] & & & \\
      &   &   &   &   &   &           & [$-2$; $-$; 0,2,8,10],
                                        [0; $-$;  0,2,4,12] & & & \\
      &   &   &   &   &   &           & [0; $-$;  0,2,6,10],
                                        [0; $-$;  0,4,6,8] & & & \\
      &   &   &   &   &   &           & [2; $-$;  0,2,4,10],
                                        [2; $-$;  0,2,6,8] & 0 & 10 & 15\\
 \hline
$q^{10}$ & 2 & 0 & 1 & 0 & 0 & $-$ & [$-$; $-$; 0,20],[$-$; $-$; 2,18],
 [$-$; $-$; 4,16]    &  &  &  \\
      &   &   &   &   &   &     & [$-$; $-$; 6,14],[$-$; $-$; 8,12]
 & 0 & 5 & \\
      & 4 & 0 & 1 & 1 & 0 & $-2\pi/M$ & [$-2$; $-$; 0,2,4,16],
                                        [$-2$; $-$; 0,2,6,14] & & & \\
      &   &   &   &   &   &           & [$-2$; $-$; 0,4,6,12],
                                        [$-2$; $-$; 2,4,6,10] & & & \\
      &   &   &   &   &   &           & [$-2$; $-$; 0,2,8,12],
                                        [$-2$; $-$; 0,4,8,10] & & & \\
      &   &   &   &   &   &           & [0; $-$;  0,2,4,14],
                                        [0; $-$;  0,2,6,12] & & & \\
      &   &   &   &   &   &           & [0; $-$;  0,4,6,10],
                                        [0; $-$;  2,4,6,8] & & & \\
      &   &   &   &   &   &           & [0; $-$;  0,2,8,10],
                                        [2; $-$;  0,2,4,12] & & & \\
      &   &   &   &   &   &           & [2; $-$;  0,2,6,10],
                                        [2; $-$;  0,4,6,8] & 0 & 14 & 19\\
 \hline
\end{tabular}
}
\end{center}
\end{table}

\begin{table}
\begin{center}
{\footnotesize
\caption{The first terms in the partition function for the sector
$Q=1$ and $m_--m_{++}=1$, where $m_+=2m_{ns}+3m_-+4m_{-2s}-2$, and the
momentum ranges are: $-\frac{\pi}{M}(m_{-2s}+m_- -3) \leq P_j^+
<\infty,\ -\frac{\pi}{M}(m_{-2s}+m_- -3) \leq P_j^{-2s} \leq
\frac{\pi}{M}(m_{-2s}+m_- -3),\ -\frac{\pi}{M}(m_{ns}+2m_{-2s}+2m_-
-3) \leq P_j^{ns} \leq \frac{\pi}{M}(m_{ns}+2m_{-2s}+2m_- -3) $.  }
\vspace{12pt}
\label{sec11}
\begin{tabular}{|l|cccc|c|c|l|r|r|} \hline
{Order} & $m_+$ & $m_{-2s}$ & $m_-$ & $m_{ns}$ & $P_{min}^{+,-2s}$
& $P_{min}^{ns}$ & $[P^{ns};P^{-2s};P^{+}]$ (Units
of $\frac{\pi}{M}$) &{$N$} & {Tot} \\ \hline\hline
$q^1$ & 1 & 0 & 1 & 0 & $2\pi/M$ & $-$ & [$-$; $-$; 2] &1&1\\ \hline
$q^2$ & 1 & 0 & 1 & 0 & $2\pi/M$ & $-$ & [$-$; $-$; 4] &1&1\\ \hline
$q^3$ & 1 & 0 & 1 & 0 & $2\pi/M$ & $-$ & [$-$; $-$; 6] &1&1\\ \hline
$q^4$ & 1 & 0 & 1 & 0 & $2\pi/M$ & $-$ & [$-$; $-$; 8] &1&1\\ \hline
$q^5$ & 1 & 0 & 1 & 0 & $2\pi/M$ & $-$ & [$-$; $-$; 10] &1&1\\ \hline
$q^6$ & 1 & 0 & 1 & 0 & $2\pi/M$ & $-$ & [$-$; $-$; 12] &1&\\
      & 3 & 0 & 1 & 1 & $2\pi/M$ & $0$ & [0; $-$; 2,4,6] & 1 & 2 \\ \hline
$q^7$ & 1 & 0 & 1 & 0 & $2\pi/M$ & $-$ & [$-$; $-$; 14] &1&\\
      & 3 & 0 & 1 & 1 & $2\pi/M$ & $0$ & [0; $-$; 2,4,8] & 1 & 2 \\ \hline
$q^8$ & 1 & 0 & 1 & 0 & $2\pi/M$ & $-$ & [$-$; $-$; 16] &1&\\
      & 3 & 0 & 1 & 1 & $2\pi/M$ & $0$ & [0; $-$; 2,4,10],
      [0; $-$; 2,6,8]  & 2 &  \\
      & 4 & 0 & 2 & 0 & $ \pi/M$ & $-$ & [$-$; $-$; 1,3,5,7] & 1 & 4
           \\ \hline
$q^9$ & 1 & 0 & 1 & 0 & $2\pi/M$ & $-$ & [$-$; $-$; 18] &1&\\
      & 3 & 0 & 1 & 1 & $2\pi/M$ & $0$ & [0; $-$; 2,4,12],
      [0; $-$; 2,6,10],[0; $-$; 4,6,8]  & 3 &  \\
      & 4 & 0 & 2 & 0 & $ \pi/M$ & $-$ & [$-$; $-$; 1,3,5,9] & 1 & 5
           \\ \hline
$q^{10}$ & 1 & 0 & 1 & 0 & $2\pi/M$ & $-$ & [$-$; $-$; 20] &1&\\
      & 3 & 0 & 1 & 1 & $2\pi/M$ & $0$ & [0; $-$; 2,4,14],
      [0; $-$; 2,6,12],[0; $-$; 4,6,10]  &  &  \\
      &   &   &   &   &          &     & [0; $-$; 2,8,10] & 4 & \\
      & 4 & 0 & 2 & 0 & $ \pi/M$ & $-$ & [$-$; $-$; 1,3,5,11],
      [$-$; $-$; 1,3,7,9] & 2 & 7
           \\ \hline
$q^{11}$ & 1 & 0 & 1 & 0 & $2\pi/M$ & $-$ & [$-$; $-$; 22] &1&\\
      & 3 & 0 & 1 & 1 & $2\pi/M$ & $0$ & [0; $-$; 2,4,16],
      [0; $-$; 2,6,14],[0; $-$; 4,6,12]  &  &  \\
      &   &   &   &   &          &     & [0; $-$; 2,8,12],
       [0; $-$; 4,8,10] & 5 & \\
      & 4 & 0 & 2 & 0 & $ \pi/M$ & $-$ & [$-$; $-$; 1,3,5,13],
      [$-$; $-$; 1,3,7,11] &  &   \\
      & & & & & & & [$-$; $-$; 1,5,7,9] & 3 & 9 \\ \hline
$q^{12}$ & 1 & 0 & 1 & 0 & $2\pi/M$ & $-$ & [$-$; $-$; 24] &1&\\
      & 3 & 0 & 1 & 1 & $2\pi/M$ & $0$ & [0; $-$; 2,4,18],
      [0; $-$; 2,6,16],[0; $-$; 4,6,14]  &  &  \\
      &   &   &   &   &          &     & [0; $-$; 2,8,14],
       [0; $-$; 4,8,12],[0; $-$; 2,10,12] &  & \\
      &   &   &   &   &          &     & [0; $-$; 6,8,10] &7 & \\
      & 4 & 0 & 2 & 0 & $ \pi/M$ & $-$ & [$-$; $-$; 1,3,5,15],
      [$-$; $-$; 1,3,7,13] &  &   \\
      & & & & & & & [$-$; $-$; 1,5,7,11],[$-$; $-$; 1,3,9,11] &  & \\
      & & & & & & & [$-$; $-$; 3,5,7,9] & 5 & 13\\ \hline
\end{tabular}}
\end{center}
\end{table}

\begin{table}
\begin{center}
{\footnotesize
\caption{The first terms in the partition function for the sector
$Q=1$ and $m_--m_{++}=-1$, where $m_+=2m_{ns}+3m_-+4m_{-2s}+2$, and
the momentum ranges are: $-\frac{\pi}{M}(m_{-2s}+m_- -1) \leq P_j^+
<\infty,\ -\frac{\pi}{M}(m_{-2s}+m_- -1) \leq P_j^{-2s} \leq
\frac{\pi}{M}(m_{-2s}+m_- -1),\ -\frac{\pi}{M}(m_{ns}+2m_{-2s}+2m_-
-1) \leq P_j^{ns} \leq \frac{\pi}{M}(m_{ns}+2m_{-2s}+2m_- -1) $.  }
\vspace{12pt}
\label{sec12}
\begin{tabular}{|l|cccc|c|c|l|r|r|} \hline
{Order} & $m_+$ & $m_{-2s}$ & $m_-$ & $m_{ns}$ & $P_{min}^{+,-2s}$
& $P_{min}^{ns}$ & $[P^{ns};P^{-2s};P^{+}]$ (Units
of $\frac{\pi}{M}$) &{$N$} & {Tot} \\ \hline\hline
$q^2$ & 2 & 0 & 0 & 0 & $\pi/M$ & $-$ & [$-$; $-$; 1,3] & 1 & 1 \\ \hline
$q^3$ & 2 & 0 & 0 & 0 &$\pi/M$ & $-$ & [$-$; $-$; 1,5] & 1 & 1 \\ \hline
$q^4$ & 2 & 0 & 0 & 0 &$\pi/M$ & $-$ & [$-$; $-$; 1,7],[$-$; $-$; 3,5] &
 2 & 2 \\ \hline
$q^5$ & 2 & 0 & 0 & 0 &$\pi/M$ & $-$ & [$-$; $-$; 1,9],[$-$; $-$; 3,7] &
 2 & 2 \\ \hline
$q^6$ & 2 & 0 & 0 & 0 &$\pi/M$ & $-$ & [$-$; $-$; 1,11],[$-$; $-$; 3,9],
 [$-$; $-$; 5,7] & 3 &  \\ \hline
$q^7$ & 2 & 0 & 0 & 0 &$\pi/M$ & $-$ & [$-$; $-$; 1,13],[$-$; $-$; 3,11],
 [$-$; $-$; 5,9] & 3 &  \\
      & 4 & 0 & 0 & 1 &$\pi/M$ & $-2\pi/M$ & [-2; $-$; 1,3,5,7] & 1 &
4 \\ \hline
$q^8$ & 2 & 0 & 0 & 0 &$\pi/M$ & $-$ & [$-$; $-$; 1,15],[$-$; $-$;3,13],
 [$-$; $-$; 5,11] & & \\
      &   &   &   &   &        &     & [$-$; $-$; 7,9] & 4 & \\
      & 4 & 0 & 0 & 1 &$\pi/M$ & $-2\pi/M$ & [-2; $-$; 1,3,5,9],
  [0,$-$,1,3,5,7] & 2 & 6 \\ \hline
$q^9$ & 2 & 0 & 0 & 0 &$\pi/M$ & $-$ & [$-$; $-$; 1,17],[$-$; $-$;3,15],
  [$-$; $-$; 5,13] & & \\
      &   &   &   &   &        &     & [$-$; $-$; 7,11] & 4 & \\
      & 4 & 0 & 0 & 1 &$\pi/M$ & $-2\pi/M$ & [-2; $-$; 1,3,5,11],
  [-2; $-$; 1,3,7,9] &  &  \\
      &   &   &   &   &  & & [0; $-$; 1,3,5,9],[2; $-$; 1,3,5,7] & 4 &
  8\\ \hline
$q^{10}$ & 2 & 0 & 0 &0 & $\pi/M$ & $-$ & [$-$; $-$; 1,19],[$-$; $-$;3,17],
  [$-$; $-$; 5,15] & & \\
      &   &   &   &   &    &     & [$-$; $-$; 7,13],[$-$; $-$; 9,11] & 5 & \\
      & 4 & 0 & 0 & 1 &$\pi/M$ & $-2\pi/M$ & [-2; $-$; 1,3,5,13],
  [-2; $-$; 1,3,7,11] &  &  \\
      &   &   &   &   &  & & [-2; $-$; 1,3,7,9],[0; $-$; 1,3,5,9] &  &\\
      &   &   &   &   &  & &[0; $-$; 1,3,7,9],[2; $-$; 1,3,5,9] &6  &\\
      & 5 & 0 & 1 & 0 & 0& $-$ & [$-$; $-$; 0,2,4,6,8] & 1 & 12\\ \hline
$q^{11}$ & 2 & 0 & 0 &0 & $\pi/M$ & $-$ & [$-$; $-$; 1,21],[$-$; $-$;3,19],
  [$-$; $-$; 5,17] & & \\
      &   &   &   &   &    &     & [$-$; $-$; 7,15],[$-$; $-$; 9,13] & 5 & \\
      & 4 & 0 & 0 & 1 &$\pi/M$ & $-2\pi/M$ & [-2; $-$; 1,3,5,15],
  [-2; $-$; 1,3,7,13] &  &  \\
     &   &   &   &   &  & & [-2; $-$; 1,5,7,11],[-2; $-$; 3,5,7,9] &  &\\
      &  &  &  &  & &  & [-2; $-$; 1,3,9,11],  [0; $-$; 1,3,5,13] &  &  \\
      &   &   &   &   &  & & [0; $-$; 1,3,7,11],[0; $-$; 1,5,7,9] &  &\\
      &   &   &   &   &  & &[2; $-$; 1,3,5,11],[2; $-$; 1,3,7,9] &10  &\\
      & 5 & 0 & 1 & 0 & 0& $-$ & [$-$; $-$; 0,2,4,6,10] & 1 & 16\\ \hline
$q^{12}$ & 2 & 0 & 0 &0 & $\pi/M$ & $-$ & [$-$; $-$; 1,23],[$-$; $-$;3,21],
  [$-$; $-$; 5,19] & & \\
      &   &   &   &   &    &     & [$-$; $-$; 7,17],[$-$; $-$; 9,15],
  [$-$; $-$; 11,13] & 6 & \\
      & 4 & 0 & 0 & 1 &$\pi/M$ & $-2\pi/M$ & [-2; $-$; 1,3,5,17],
  [-2; $-$; 1,3,7,15] &  &  \\
     &   &   &   &   &  & & [-2; $-$; 1,5,7,13],[-2; $-$; 3,5,7,11] &  &\\
      &  &  &  &  & &  & [-2; $-$; 1,3,9,13],  [-2; $-$; 1,5,9,11] & &  \\
      &   &   &   &   &  & & [0; $-$; 1,3,5,15],[0; $-$; 1,3,7,13] &  &\\
      &   &   &   &   &  & &[0; $-$; 1,5,7,11 ],[0; $-$; 3,5,7,9] &  &\\
      &   &   &   &   &  & &[0; $-$; 1,3,9,11 ],[2; $-$; 1,3,5,13] &  &\\
      &   &   &   &   &  & &[2; $-$; 1,3,7,11],[2; $-$; 1,5,7,9] & 14 &\\
      & 5 & 0 & 1 & 0 & 0& $-$ & [$-$; $-$; 0,2,4,6,12],
 [$-$; $-$; 0,2,4,8,10] & 2 & 22\\ \hline
\end{tabular}}
\end{center}
\end{table}

\begin{table}
\begin{center}
{\footnotesize
\caption{The first few terms in the sector of the partition function
for the sector $Q=1$ and $m_-=m_{++}=0$, where
$m_+=2m_{ns}+3m_-+4m_{-2s}$, and the momentum ranges are:
$-\frac{\pi}{M}(m_{-2s}+m_- -2) \leq P_j^+ <\infty,\
-\frac{\pi}{M}(m_{-2s}+m_- -2) \leq P_j^{-2s} \leq
\frac{\pi}{M}(m_{-2s}+m_- -2),\ -\frac{\pi}{M}(m_{ns}+2m_{-2s}+2m_-
-1) \leq P_j^{ns} \leq \frac{\pi}{M}(m_{ns}+2m_{-2s}+2m_- -1) $. }
\label{sec13}
\vspace{12pt}
\begin{tabular}{|l|cccc|c|c|l|r|r|} \hline
{Order} & $m_+$ & $m_{-2s}$ & $m_-$ & $m_{ns}$ & $P_{min}^{+,-2s}$
& $P_{min}^{ns}$ & $[P^{ns};P^{-2s};P^{+}]$ (Units
of $\frac{\pi}{M}$) &{$N$} & {Tot} \\ \hline\hline
$q^0$ & 0 & 0 & 0 & 0 & $-$ & $-$ & [$-$; $-$; $-$] & 1 & 1 \\ \hline
$q^3$ & 2 & 0 & 0 & 1 & $2\pi/M$ & 0 & [0; $-$; 2,4] & 1 & 1 \\ \hline
$q^4$ & 2 & 0 & 0 & 1 & $2\pi/M$ & 0 & [0; $-$; 2,6] & 1 & 1 \\ \hline
$q^5$ & 2 & 0 & 0 & 1 & $2\pi/M$ & 0 & [0; $-$; 2,8],[0; $-$; 4,6]
  & 2 & 2 \\ \hline
$q^6$ & 2 & 0 & 0 & 1 & $2\pi/M$ & 0 & [0; $-$; 2,10],[0; $-$; 4,8]
  & 2 & 2 \\ \hline
$q^7$ & 2 & 0 & 0 & 1 & $2\pi/M$ & 0 & [0; $-$; 2,12],[0; $-$; 4,10],
 [0; $-$; 6,8]  & 3 & 3 \\ \hline
$q^8$ & 2 & 0 & 0 & 1 & $2\pi/M$ & 0 & [0; $-$; 2,14],[0; $-$; 4,12],
 [0; $-$; 6,10]  & 3 & 3 \\ \hline
$q^9$ & 2 & 0 & 0 & 1 & $2\pi/M$ & 0 & [0; $-$; 2,16],[0; $-$; 4,14],
 [0; $-$; 6,12]  &  &  \\
      &   &   &   &   &          &   & [0; $-$; 8,10] & 4 & 4 \\ \hline
$q^{10}$ & 2 & 0 & 0 & 1 & $2\pi/M$ & 0 & [0; $-$; 2,18],[0; $-$; 4,16],
 [0; $-$; 6,14]  &  &  \\
      &   &   &   &   &          &   & [0; $-$; 8,12] & 4 &  \\
      & 4 & 0 & 0 & 2 & $2\pi/M$ & $-\pi/M$ & [-1,1; $-$; 2,4,6,8] & 1
& 5 \\ \hline
$q^{11}$ & 2 & 0 & 0 & 1 & $2\pi/M$ & 0 & [0; $-$; 2,20],[0; $-$; 4,18],
 [0; $-$; 6,16]  &  &  \\
      &   &   &   &   &          &   & [0; $-$; 8,14],[0; $-$; 10,12] & 5 &  \\
      & 4 & 0 & 0 & 2 & $2\pi/M$ & $-\pi/M$ & [-1,1; $-$; 2,4,6,10] & 1
& 6 \\ \hline
$q^{12}$ & 2 & 0 & 0 & 1 & $2\pi/M$ & 0 & [0; $-$; 2,22],[0; $-$; 4,20],
 [0; $-$; 6,18]  &  &  \\
      &   &   &   &   &          &   & [0; $-$; 8,16],[0; $-$; 10,14] & 5 &  \\
      & 4 & 0 & 0 & 2 & $2\pi/M$ & $-\pi/M$ & [-1,1; $-$; 2,4,6,12],
    [-1,1; $-$; 2,4,8,10]& 2& 7 \\ \hline
\end{tabular}}
\end{center}
\end{table}

\begin{table}
\begin{center}
{\footnotesize
\caption{The first few terms for the sector of the partition function
corresponding to $Q=1$ and $m_-=m_{++}\neq0$, where
$m_+=2m_{ns}+3m_-+4m_{-2s}$, and the
momentum ranges are:
$-\frac{\pi}{M}(m_{-2s}+m_- -3) \leq P_j^+
<\infty,$
$-\frac{\pi}{M}(m_{-2s}+m_- -1) \leq P_j^{-2s}
  \leq \frac{\pi}{M}(m_{-2s}+m_- -3),$
$-\frac{\pi}{M}(m_{ns}+2m_{-2s}+2m_- ) \leq P_j^{ns}
  \leq \frac{\pi}{M}(m_{ns}+2m_{-2s}+2m_- -2)
$, and  there is a shift of $-\frac{2\pi}{M} (\half
m_{ns}+m_{-2s}+m_-)$. }
\label{sec14}
\vspace{12pt}
\begin{tabular}{|l|cccc|c|c|l|c|r|r|} \hline
{ Order} & $m_+$ & $m_{-2s}$ & $m_-$ & $m_{ns}$ & $P_{min}^{+,-2s}$
& $P_{min}^{ns}$ & $[P^{ns};P^{-2s};P^{+}]$ (Units
of $\frac{\pi}{M}$) &{ Shift} & { N} & { Tot} \\ \hline\hline
$q^5$ & 3 & 0 & 1&0 & $2\pi/M$ & $-$ & [$-$; $-$; 2,4,6] & $-2\pi/M$ & 1 & 1
\\ \hline
$q^6$ & 3 & 0 & 1&0 & $2\pi/M$ & $-$ & [$-$; $-$; 2,4,8] & $-2\pi/M$ & 1 & 1
\\ \hline
$q^7$ & 3 & 0 & 1&0 & $2\pi/M$ & $-$ & [$-$; $-$; 2,4,10],
 [$-$; $-$; 2,6,8] & $-2\pi/M$ & 2 & 2 \\ \hline
$q^8$ & 3 & 0 & 1&0 & $2\pi/M$ & $-$ & [$-$; $-$; 2,4,12],
 [$-$; $-$; 2,6,10] &  & &  \\
      & & & & & & & [$-$; $-$; 4,6,8] & $-2\pi/M$ & 3 & 3 \\ \hline
$q^9$ & 3 & 0 & 1&0 & $2\pi/M$ & $-$ & [$-$; $-$; 2,4,14],
 [$-$; $-$; 2,6,12] &  & &  \\
      & & & & & & & [$-$; $-$; 4,6,10],[$-$; $-$; 2,8,10] & $-2\pi/M$
& 4 & 4 \\ \hline
$q^{10}$ & 3 & 0 & 1&0 & $2\pi/M$ & $-$ & [$-$; $-$; 2,4,16],
 [$-$; $-$; 2,6,14] &  & &  \\
      & & & & & & & [$-$; $-$; 4,6,12],[$-$; $-$; 2,8,12]&  & &  \\
      & & & & & & & [$-$; $-$; 4,8,10] &$-2\pi/M$  &5 & 5 \\ \hline
$q^{11}$ & 3 & 0 & 1&0 & $2\pi/M$ & $-$ & [$-$; $-$; 2,4,18],
 [$-$; $-$; 2,6,16] &  & &  \\
      & & & & & & & [$-$; $-$; 4,6,14],[$-$; $-$; 2,8,14]&  & &  \\
      & & & & & & & [$-$; $-$; 4,8,12],[$-$; $-$; 6,8,10] &  & &  \\
      & & & & & & & [$-$; $-$; 2,10,12] & $-2\pi/M$ &7 &7  \\ \hline
$q^{12}$ & 3 & 0 & 1&0 & $2\pi/M$ & $-$ & [$-$; $-$; 2,4,20],
 [$-$; $-$; 2,6,18] &  & &  \\
      & & & & & & & [$-$; $-$; 4,6,16],[$-$; $-$; 2,8,16]&  & &  \\
      & & & & & & & [$-$; $-$; 4,8,14],[$-$; $-$; 6,8,12] &  & &  \\
      & & & & & & & [$-$; $-$; 2,10,12],[$-$; $-$; 4,10,12] &$-2\pi/M$ &8 &  \\
      &5&0&1&1&$2\pi/M$&$-3\pi/M$&[$-3$; $-$; 2,4,6,8,10]&$-3\pi/M$& 1 & 9\\
\hline
\end{tabular}}
\end{center}
\end{table}

\begin{table}
\begin{center}
{\footnotesize
\caption{The first few terms for the sector of the partition function
corresponding to $Q=1$ and $m_-=m_{++}\neq0$, where
$m_+=2m_{ns}+3m_-+4m_{-2s}$, and the momentum ranges are:
$-\frac{\pi}{M}(m_{-2s}+m_- -1) \leq P_j^+ <\infty,\
-\frac{\pi}{M}(m_{-2s}+m_- -1) \leq P_j^{-2s} \leq
\frac{\pi}{M}(m_{-2s}+m_- -3),\ -\frac{\pi}{M}(m_{ns}+2m_{-2s}+2m_-
-2) \leq P_j^{ns} \leq \frac{\pi}{M}(m_{ns}+2m_{-2s}+2m_- ) $. There
is an additive shift in the total momentum of $\frac{2\pi}{M} (\half
m_{ns}+m_{-2s}+m_-)$. }
\vspace{12pt}
\label{sec15}
\begin{tabular}{|l|cccc|c|c|l|c|r|r|} \hline
{ Order} & $m_+$ & $m_{-2s}$ & $m_-$ & $m_{ns}$ & $P_{min}^{+,-2s}$
& $P_{min}^{ns}$ & $[P^{ns};P^{-2s};P^{+}]$ (Units
of $\frac{\pi}{M}$) &{Shift} & {$N$} & {Tot} \\ \hline\hline
$q^4$ & 3 & 0 & 1&0 & $0$ & $-$ & [$-$; $-$; 0,2,4] & $2\pi/M$ & 1 & 1
\\ \hline
$q^5$ & 3 & 0 & 1&0 & $0$ & $-$ & [$-$; $-$; 0,2,6] & $2\pi/M$ & 1 & 1
\\ \hline
$q^6$ & 3 & 0 & 1&0 & $0$ & $-$ & [$-$; $-$; 0,2,8],
 [$-$; $-$; 0,4,6] & $2\pi/M$ & 2 & 2 \\ \hline
$q^7$ & 3 & 0 & 1&0 & $0$ & $-$ & [$-$; $-$; 0,2,10],
 [$-$; $-$; 0,4,8] &  & &  \\
      & & & & & & & [$-$; $-$; 2,4,6] & $2\pi/M$ & 3 & 3 \\ \hline
$q^8$ & 3 & 0 & 1&0 & $0$ & $-$ & [$-$; $-$; 0,2,12],
 [$-$; $-$; 0,4,10] &  & &  \\
      & & & & & & & [$-$; $-$; 2,4,8],[$-$; $-$; 0,6,8] & $2\pi/M$ & 4 & 4 \\
\hline
$q^9$ & 3 & 0 & 1&0 & $0$ & $-$ & [$-$; $-$; 0,2,14],
 [$-$; $-$; 0,4,12] &  & &  \\
      & & & & & & & [$-$; $-$; 2,4,10],[$-$; $-$; 0,6,10]&  & &  \\
      & & & & & & & [$-$; $-$; 2,6,8] &$2\pi/M$  &5 & 5 \\ \hline
$q^{10}$ & 3 & 0 & 1&0 & $0$ & $-$ & [$-$; $-$; 0,2,16],
 [$-$; $-$; 0,4,14] &  & &  \\
      & & & & & & & [$-$; $-$; 2,4,12],[$-$; $-$; 0,6,12]&  & &  \\
      & & & & & & & [$-$; $-$; 2,6,10],[$-$; $-$; 4,6,8] &  & &  \\
      & & & & & & & [$-$; $-$; 0,8,10] & $2\pi/M$ &7 &7  \\ \hline
$q^{11}$ & 3 & 0 & &0 & $0$ & $-$ & [$-$; $-$; 0,2,18],
 [$-$; $-$; 0,4,16] &  & &  \\
      & & & & & & & [$-$; $-$; 2,4,14],[$-$; $-$; 0,6,14]&  & &  \\
      & & & & & & & [$-$; $-$; 2,6,12],[$-$; $-$; 4,6,10] &  & &  \\
      & & & & & & & [$-$; $-$; 0,8,12],[$-$; $-$; 2,8,10] &$2\pi/M$ &8 &  \\
      &5&0&1&1&$0$&$-\pi/M$&[$-1$; $-$; 0,2,4,6,8]&$3\pi/M$& 1 & 9\\
\hline
$q^{12}$ & 3 & 0 & 1&0 & $0$ & $-$ & [$-$; $-$; 0,2,20],
 [$-$; $-$; 0,4,18] &  & &  \\
      & & & & & & & [$-$; $-$; 2,4,16],[$-$; $-$; 0,6,16]&  & &  \\
      & & & & & & & [$-$; $-$; 2,6,14],[$-$; $-$; 4,6,12] &  & &  \\
      & & & & & & & [$-$; $-$; 0,8,14],[$-$; $-$; 2,8,12] & & &  \\
      & & & & & & & [$-$; $-$; 0,10,12],[$-$; $-$; 4,8,10] &$2\pi/M$ &10 &  \\
      &5&0&1&1&$0$&$-\pi/M$&[$-1$; $-$; 0,2,4,6,10],& & &\\
& & & & & & & [1; $-$; 0,2,4,6,8]&$3\pi/M$& 2 & 12\\ \hline
\end{tabular}}
\end{center}
\end{table}


\baselineskip12pt
\begin{thebibliography}{99}
\bibitem{temperley71} H.N.V. Temperley and E.H. Lieb,  Proc. Roy. Soc.
London A322 (1971) 251.
\bibitem{baxter73} R.J. Baxter, J. Phys. C6 (1973) L445.
\bibitem{baxter82} R.J. Baxter, {\it Exactly solved models in
statistical mechanics} (Academic Press, London, 1982).
\bibitem{kedem93} R. Kedem and B.M. McCoy, J. Stat. Phys. (in press),
hep-th/9210129.
\bibitem{albertini92}G. Albertini, S. Dasmahapatra and B. M. McCoy, Phys.
Lett. A170 (1992) 397.
\bibitem{baxter82b} R.J. Baxter and P.A. Pearce, J. Phys. A15 (1982) 897.
\bibitem{bazhanov89}V.V. Bazhanov and N.Yu. Reshetikhin, Int. J.
Mod. Phys. A4 (1989) 115.
\bibitem{albertini89} G. Albertini, B.M. McCoy and J.H.H. Perk, Phys.
Lett. A 135 (1989) 159, and
in {\it Advanced Studies in Pure Mathematics}  19
ed. M. Jimbo, T. Miwa and A. Tsuchiya (Kinokuniya-Academic, Tokyo, 1989)
p. 1.
\bibitem{bazhanov90} V.V. Bazhanov and Yu.G. Stroganov,
 J. Stat. Phys. 59 (1990) 799.
\bibitem{baxter90}R.J. Baxter, V.V. Bazhanov and J.H.H. Perk, Int. J. Mod.
Phys. B4 (1990) 803.
\bibitem{albertini92b}G. Albertini, J. Phys. A25 (1992) 1799.
\bibitem{pearce92} P.A. Pearce,
Int. J. Mod. Phys. A7, Suppl.1B (1992) 791.
\bibitem{lepowsky85}J. Lepowsky and M. Primc, {\it Structure of the standard
modules for the affine Lie algebra $A_1^{(1)}$}, Contemporary
Mathematics, Vol. 46 (AMS, Providence, 1985).
\bibitem{rocha85}A. Rocha-Caridi, in {\it Vertex Operators in Mathematics and
Physics}, ed. J. Lepowsky, S. Mandelstam and I.M. Singer
(Springer, Berlin, 1985) p. 451.
\bibitem{dotsenko84}Vl.S. Dotsenko, Nucl. Phys. B235[FS11] (1984) 54.
\bibitem{zamolodchikov85}A.B. Zamolodchikov and V.A. Fateev, Sov. Phys. JETP
62 (1985) 215.
\bibitem{cardy86a} J.L. Cardy, Nucl. Phys. B275[FS17] (1986) 200.
\bibitem{kedem93c} R. Kedem, T.R. Klassen, B.M. McCoy, and E. Melzer,
 Phys. Lett. B (in press), hep-th/9301046.
\bibitem{kedem93b} R. Kedem, J. Stat. Phys. (in press), hep-th/9210146.
\bibitem{gepner87} D. Gepner and Z. Qiu, Nucl. Phys. B285 [FS19] (1987) 423.
\bibitem{albertini92a} G. Albertini, S. Dasmahapatra and B.M. McCoy, Int. J.
Mod. Phys. A7, Suppl. 1A (1992) 1.
\bibitem{klumper91} A. Kl\"umper and P.A. Pearce, J. Stat. Phys. 64 (1991)
13; Physica A 183 (1992) 304.
\bibitem{stanley72} R.P. Stanley, {\it Ordered structures and partitions}, Mem.
 Amer. Math. Soc. 119 (1972).
\bibitem{andrews76} G.E. Andrews, {\it The Theory of Partitions}
  (Addison-Wesley, London, 1976).
\bibitem{kac84}V.G. Kac and D.H. Peterson, Adv. in Math. 53 (1984) 125.
\bibitem{jimbo84} M. Jimbo and T. Miwa, Adv. Stud. in Pure Math. 4 (1984)
97.
\bibitem{goddard86} P. Goddard, A. Kent and D. Olive, Commun. Math.
 Phys. 103 (1986) 105.
\bibitem{feigin83}B.L. Feigin and D.B. Fuchs, Funct. Anal. Appl. 17 (1983)
241.
\bibitem{felder89}G. Felder, Nucl. Phys. B317 (1989) 215.
\bibitem{belavin84}A.A. Belavin, A.M. Polyakov and A.B. Zamolodchikov, J.
Stat. Phys. 34 (1984) 763; Nucl. Phys. B241 (1984) 337.
\bibitem{fateev87}V.A. Fateev and A.B. Zamolodchikov, Nucl. Phys.
B280[FS18] (1987) 644.
\bibitem{altschuler90}D. Altschuler, M. Bauer and H. Saleur, J. Phys. A23
(1990) L789.
\bibitem{distler90} J. Distler and Z. Qiu, Nucl. Phys. B336 (1990) 533.
\bibitem{kedem93a} R. Kedem, T.R. Klassen, B.M. McCoy and E. Melzer,
 Phys. Lett. B (in press), hep-th/9211102.
\bibitem{zamolodchikov89} A.B. Zamolodchikov, in
{\it Advanced Studies in Pure Mathematics}  19
ed. M. Jimbo, T. Miwa and A. Tsuchiya (Kinokuniya-Academic,
 Tokyo, 1989) p. 641.
\bibitem{zamolodchikov88} A.B. Zamolodchikov, Int. J. Mod. Phys. A3 (1988)
743.
\bibitem{koberle79} R. K\"oberle and J.A. Swieca, Phys. Lett. B86 (1979) 209.
\bibitem{baxter80} R.J. Baxter, J. Phys. A13 (1980) L61; J. Stat. Phys.
26 (1981) 427.
\bibitem{kashiwara86} Kashiwara and Miwa, Nucl. Phys. B275 (1986) 121.
\bibitem{pasquier87} V. Pasquier, J. Phys. A20 (1987) L217 and L221.
\bibitem{andrews84} G.E. Andrews, R.J. Baxter and P.J. Forrester,
J. Stat. Phys.  35 (1984) 193.
\bibitem{fendley89} P. Fendley and P. Ginsparg, Nucl. Phys. B324 (1989) 549.
\bibitem{zamolodchikov87} A.B. Zamolodchikov, Sov. J. Nucl. Phys. 46 (1987)
1090.
\bibitem{ludwig87}A. Ludwig and J.L. Cardy, Nucl. Phys. B285 [FS19] (1987)
687.
\bibitem{zamolodchikov89a} A.B. Zamolodchikov,
Landau Institute preprint (1989).
\bibitem{bernard90} D. Bernard and A. LeClair, Nucl. Phys. B340 (1990) 721.
\bibitem{reshetikhin90}N.Yu. Reshetikhin and F.A. Smirnov,
 Commun. Math. Phys. 131 (1990) 157.
\bibitem{alzam91} Al.B. Zamolodchikov, Nucl. Phys. B358 (1991) 497 and 524.
\bibitem{fateev91} V.A. Fateev and Al.B. Zamolodchikov, Phys. Lett. B271
(1991) 91.
\bibitem{klassen92}T.R. Klassen and E. Melzer, Nucl. Phys. B370 (1992) 511.
\bibitem{ravanini92} F. Ravanini, Phys. Lett. B274 (1992) 345.
\bibitem{klassen91}T.R. Klassen and E. Melzer, Nucl. Phys. B
(in press), hep-th/9110047.
\bibitem{bazhanov90a}V.V. Bazhanov and N.Yu. Reshetikhin, Progr. Theor.
 Phys. Suppl. 102 (1990) 301.
\bibitem{cardy88}J.L. Cardy, in {\it Fields, strings, and critical
 phenomena}, Les Houches 1988, ed. E. Br\'ezin and J. Zinn-Justin
(North-Holland, Amsterdam, 1989).

\end{thebibliography}
\end{document}